\documentclass[aps,prl,reprint,showpacs,superscriptaddress,longbibliography]{revtex4-2}
\usepackage{mathtools,amssymb,graphicx,units}
\usepackage[usenames,dvipsnames]{color}
\usepackage[plainpages=false,pdfpagelabels,colorlinks=true,linkcolor=blue,urlcolor=magenta,citecolor=magenta,pdftitle={Title},pdfauthor={},pdfdisplaydoctitle=true,pdfduplex=DuplexFlipLongEdge]{hyperref}
\usepackage{subfigure}
\usepackage{grffile}
\usepackage{gensymb}
\usepackage{bm}
\usepackage[version=4]{mhchem}
\usepackage{color}
\usepackage{hyperref}
\usepackage{bibentry}

\usepackage{soul} 

\makeatother

\begin{document}
\title{Impact of crystal symmetries and Weyl nodes on high-harmonic generation in Weyl semimetal \texorpdfstring{TaAs}{}}
\author{Xiao Zhang}
\email{x.zhang@ifw-dresden.de}
\affiliation{Institute for Theoretical Solid State Physics, Leibniz IFW Dresden, Helmholtzstrasse 20, 01069 Dresden, Germany}
\affiliation{Institute of Theoretical Physics and W{\"u}rzburg-Dresden  Cluster of Excellence {\it ct.qmat}, Technische Universit{\"a}t Dresden, 01062 Dresden, Germany} 
\author{Jeroen van den Brink}
\email{j.van.den.brink@ifw-dresden.de}
\affiliation{Institute for Theoretical Solid State Physics, Leibniz IFW Dresden, Helmholtzstrasse 20, 01069 Dresden, Germany}
\affiliation{Institute of Theoretical Physics and W{\"u}rzburg-Dresden  Cluster of Excellence {\it ct.qmat}, Technische Universit{\"a}t Dresden, 01062 Dresden, Germany} 
\author{Jinbin Li}
\email{jinbin@pks.mpg.de}
\affiliation{Max-Planck-Institut f{\"u}r Physik komplexer Systeme, N{\"o}thnitzerstrasse 38, 01187 Dresden, Germany}
\affiliation{School of Nuclear Science and Technology, Lanzhou University, Lanzhou 730000, China}

\date{\today}

\begin{abstract}
High-harmonic generation (HHG) offers an all-optical approach to discern structural symmetries through its selection rules and probe topological phases with its spectral signatures. Here we develop a universal theoretical framework --- the Jones matrix formalism --- establishing the fundamental relationship between pulse-crystal shared symmetries and HHG selection rules. 
Applying this to the Weyl semimetal (WSM) material TaAs, shows that the anomalous harmonics excited by linearly and circularly polarized pulses are governed respectively by the shared twofold and fourfold rotational symmetries of laser pulses and lattice, rather than the topology of Weyl nodes. The common observables of HHG, including intensity, circular dichroism, ellipticity dependence, and carrier-envelope phase dependence, are not found to carry a signature of the Weyl cones. This insight into TaAs can be extended to other WSMs, 
indicating that HHG is not a particularly effective tool for investigating the topological features of WSMs. However, the Jones matrix formalism lays the groundwork for both HHG probing crystal symmetry and controlling harmonic polarization states.
\end{abstract}

\maketitle

\textit{Introduction.}
Weyl semimetals (WSMs) feature topologically protected linear band crossings known as Weyl points (WPs) in the momentum space \cite{Bauer_2007,PhysRevB.83.205101,PhysRevB.84.235126,Yan.annurev}. 
Isolated WPs, with right- or left-handed chirality (or topological charge, $\chi=\pm 1$), act as monopoles for Berry curvature, 
requiring the material to break either spatial inversion ($\mathcal{I}$) or time-reversal symmetry ($\mathcal{T}$) for their emergence \cite{RevModPhys.82.1959,RevModPhys.82.1539,BernevigJPSJ2018,Klaus2016}. 
These unusual electronic features give rise to appealing nonlinear optical effects, like giant photocurrent effects and second-harmonic generation \cite{Wu2017NLO,PhysRevLett.123.246602,PhysRevB.97.085201,Juan2017}. 
Recently, the exploration of nonlinear optical effects in WSMs has been extended to the realm of strong field physics such as high-harmonic generation (HHG). 
Experimentally, both odd and even harmonics have been observed in $\beta$-WP$_2$ crystals \cite{Lv2021HHG}, a type-II WSM characterized by broken $\mathcal{I}$. 
Theoretical studies have shown the emergence of anomalous odd harmonics polarized perpendicular to the incident laser field within $\mathcal{T}$-breaking WSMs \cite{PhysRevB.105.155140,PhysRevA.106.033107,PhysRevB.107.224308}.
However, the mechanism behind the generation of anomalous odd and even harmonics in WSMs is still under debate.

\begin{figure*}
	\centering
	\includegraphics[width=2\columnwidth]{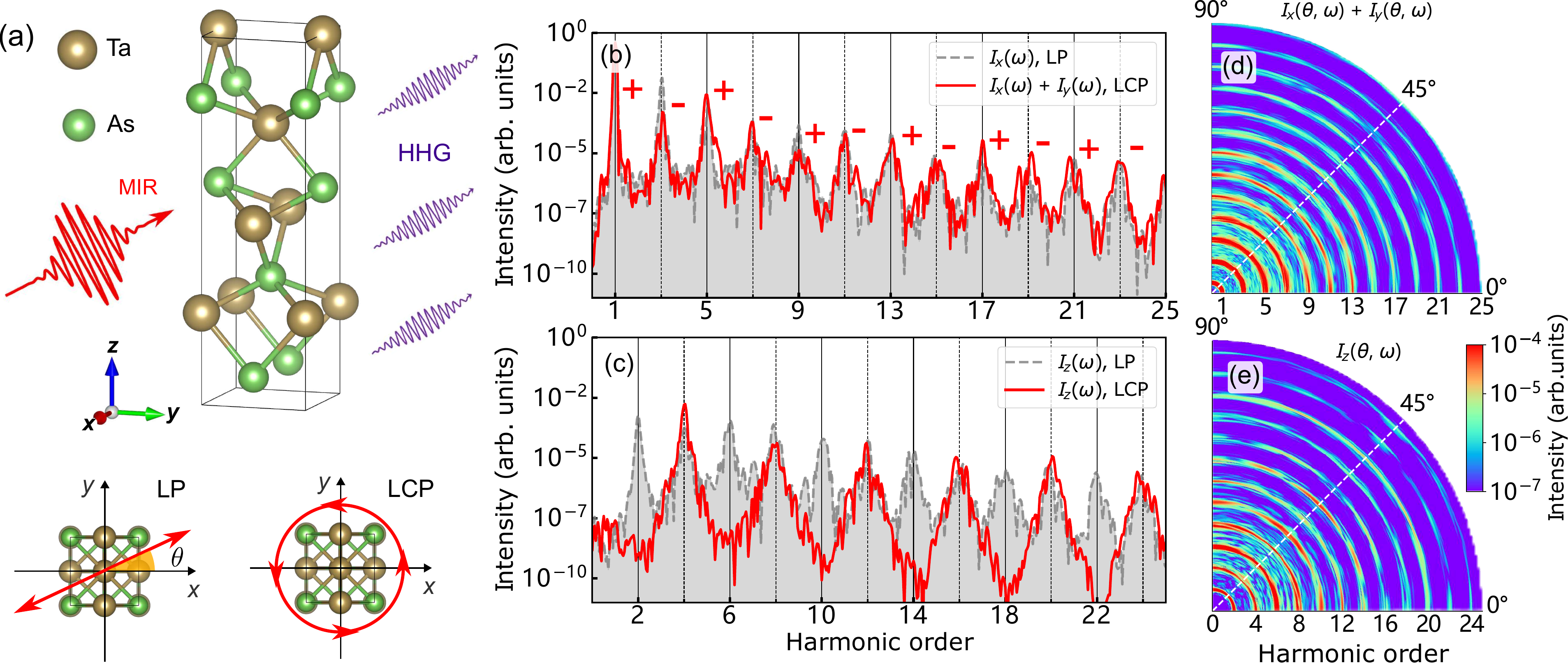}
	\caption{(a) Sketch of harmonic signals generated by MIR laser interacting with TaAs crystal, where $x$, $y$, and $z$ axes are parallel to the $[100]$, $[010]$, and $[001]$ directions, respectively. In inset $\theta$ is the angle of LP field relative to the $x$ axis. Harmonic spectra in (b) $x$-$y$ plane [$(001)$ lattice plane] and (c) $z$ direction obtained by a LP field (gray dashed curves) and by a left-handed circularly polarized driving field (red curves). The plus/minus ($+$/$-$) sign indicates harmonic helicity that is the same as/opposite to the helicity of the driving field. Orientation dependence of (d) parallel and (e) transverse harmonics excited by a LP pulse. The laser parameters are intensity $I_{0} = 5.04 \times 10^{10}$ W/cm$^2$ ($E_{0} = 0.0012$ a.u.), wavelength $\lambda=4200$ nm ($\hbar\omega=0.295$ eV), and duration $n_{\mathrm{cyc}}=20$ cycles.
	}
	\label{fig1}
\end{figure*}

Research over the past decade suggests that HHG holds promise as a potential tool for probing the crystal symmetries \cite{FLanger2017,Luu2018,PhysRevB.101.144304} and topological properties 
\cite{PhysRevLett.131.223802,PhysRevLett.120.177401,Heide2022,PhysRevX.12.021030} of solid-state materials. 
Observations across a wide range of materials indicate that the mirror and rotational symmetries of lattices are not only reflected in the anisotropic harmonic emissions \cite{You2016,LiuH2016,PhysRevLett.118.087403,Tancogne-Dejean2017} but also in the HHG selection rules \cite{Klemke2019,Zurron-Cifuentes,NariyukiSaito17}.
In Chern insulators, the helicity and circular dichroism (CD) of the harmonics serve as indicators for characterizing topological phase transitions \cite{Silva2019,Alexis2020}. 
For three dimensional topological insulators (3D TIs), the gapless surface states (Dirac cones) and gapped bulk states can be differentiated by the sensitivity of HHG to the carrier-envelope phase (CEP) \cite{Schmid2021} and the HHG ellipticity dependence \cite{PhysRevA.103.023101,Baykusheva2021nanoletter,BaiYa2021}. 
However, it remains unclear whether the crystal symmetry, the distinctive gapless Weyl cones, and the chiral Berry curvature monopoles can leave unique imprints on the high-harmonic (HH) spectra of WSMs.
Moreover, a first-principles investigation of the HHG in WSMs, considering the realistic band structure of materials, has not been reported so far.

In this letter, we address two crucial issues: i) What is the intrinsic connection between anomalous harmonics, HHG selection rules, and crystal symmetries? ii) Can HHG be effectively employed to probe Weyl cones in WSMs? We study HHG in TaAs from first principles due to its type-I Weyl cones, which are close to the Fermi energy, making it conducive for Weyl fermionic excitations in the mid-infrared (MIR) spectral range. By developing the Jones matrix formalism, we provide an analytical expression for HHG selection rules, linking the harmonic orders, polarization states, and crystal symmetries.
For the second issue, we explore the possible observables of HHG which may reflect the gapless excitations and chiral topological charges of Weyl cones. 
We find that the anomalous even harmonics are governed by the crystal symmetries rather than the topology of WPs.
The HHG CD does not reflect the chirality of topological charge because the mirror symmetries eliminate the overall chirality of TaAs.
The occurrence of anomalous HHG ellipticity dependence and sensitive HHG CEP dependence depends more on the values of laser parameters than on the presence of Weyl cones.
Blocking Weyl fermionic excitation by tuning chemical potential, the yield of each harmonic is altered slightly and does not exhibit a consistent trend of change. 
These results imply that all the aforementioned HHG signatures generally do not probe the topological properties of Weyl semimetals in an efficient manner.

Numerically, we employ a method combining time dependent Schr{\"o}dinger equation with a first-principles model \cite{wannier_SBEs_yupeng,wannier_gauge}, to calculate HHG in TaAs. 
Density functional theory (DFT) calculations, for example, 
band structure and density of states, are performed using Full-Potential Local-Orbital code \cite{PhysRevB.59.1743,fplo}. 
The first-principles model is obtained by fitting 
DFT band structure employing highly localized Wannier functions implemented in \textsc{pyfplo} module \cite{fplo}.  
Details can be found in Supplemental Material (SM) \cite{supp}.
Atomic units are used throughout this work unless indicated otherwise.

\textit{Jones matrix formalism.} 
The studies on the selection rules of solid HHG predominantly rely on experimental induction, numerical simulation, and phenomenological analysis \cite{PhysRevLett.131.223802,Klemke2019,NariyukiSaito17}, lacking a systematic framework. To fill this void, we develop the Jones matrix method (see the derivation in SM \cite{supp}). 
It shows that the harmonic field $\boldsymbol{J}(n_{H}\omega)$ can always be expressed as the product of Jones matrix and Jones vector by taking into account the space group operations in crystals
\begin{align}
    \boldsymbol{J}(n_{H}\omega)=&\mathcal{J}(n_{H})\boldsymbol{J}^{\mathrm{ir}}(n_{H}\omega),
\end{align}
where $n_{H}$ is the harmonic order, $\omega$ is the laser frequency,
and the Jones vector $\boldsymbol{J}^{\mathrm{ir}}(n_{H}\omega)$  also represents the irreducible part of the harmonic field.
The Jones matrix, expressed as
\begin{align}
    \mathcal{J}(n_{H})=\sum_{l'=1}^{N'}(\mathcal{R}')^{l'}\sum_{l=1}^{N}\exp(-i2\pi n_{H}\frac{l}{N})(\mathcal{R})^{l}, \label{jones}
\end{align}
not only reveals how the crystal symmetry controls the allowed harmonic orders and polarization states but also clarifies the underlying principles of utilizing HHG selection rules to probe crystal symmetry.
In Eq.~\eqref{jones}, $\mathcal{R}$ and $\mathcal{R}'$ are the shared symmetry matrices of external field and crystal structure, $N$ ($N'$) is the order of symmetry element $\mathcal{R}$ ($\mathcal{R}'$), i.e.,  $\mathcal{R}^{N}=E$ [$(\mathcal{R}')^{N}=E$]. $E$ is the identity element. Note that $\mathcal{R}$ and $\mathcal{R}'$ are different. 
The external field, when subjected to the operation of $\mathcal{R}$, is equivalent to being translated by $T/N$ in time (where $T$ represents the laser cycle), while the external field remains invariant under the operation of $\mathcal{R}'$. 

\textit{HHG in TaAs.}
TaAs is a Weyl semimetal crystallizing in noncentrosymmetric ($\mathcal{I}$ is broken) $I4_{1}md$ (No.109) structure \cite{YangLX2015,HuangSM2015}, see Fig.~\ref{fig1}(a).
It possesses twofold rotation $\mathcal{C}_{2(z)}$, fourfold rotation $\mathcal{C}_{4(z)}$, $\mathcal{C}_{4-(z)}$, and mirror $\mathcal{M}_{x}$, $\mathcal{M}_{y}$, $\mathcal{M}_{xy}$, $\mathcal{M}_{-xy}$ symmetries.

We begin by examining harmonic responses in TaAs driven by a linearly polarized (LP) field  along the $x$ direction.
As shown in Figs.~\ref{fig1}(b)-(c) (gray dashed curves), typical odd harmonics are observed in the $x$ (parallel) direction, 
whereas clear even-structure harmonics manifest in the $z$ (transverse) direction.
It is noteworthy that no harmonic response is observed in the $y$ direction.
Upon applying a left-handed circularly polarized (LCP) driver, depicted by the red curves, the circularly polarized odd harmonics emerge with alternating ellipticity ($\pm 1$). Along the $z$ direction, the $(4k-2)$th harmonics disappear while the ($4k$)th harmonics survive. Moreover, the ($4k$)th harmonics retain linear polarization.

Parallel harmonics are commonly referred to as ordinary, while transverse harmonics are termed anomalous \cite{PhysRevLett.124.153204,PhysRevLett.130.166903}. 
The pervious studies attribute the generation of anomalous harmonics to the influence of WPs in WSMs \cite{PhysRevB.105.155140,PhysRevA.106.033107}. 
Nonetheless, we think that crystal symmetries are the more fundamental factors resulting in the anomalous harmonics.
By using the Jones matrix method [Eq.~\eqref{jones}], all the observations in Figs.~\ref{fig1}(b)-(e) can be interpreted. As the LP field is oriented within the $x$-$y$ plane of TaAs, the shared symmetry $\mathcal{R}$ is $\mathcal{C}_{2(z)}$, see inset of Fig.~\ref{fig1}(a). 
Due to the absence of $\mathcal{M}_z$ (broken $\mathcal{I}$) in TaAs, there is no additional crystal symmetry that can maintain the LP field invariant when it deviates from the high symmetry axes. Thus, the shared symmetry $\mathcal{R}'$ equals $E$. 
Plugging $\mathcal{R}$ and $\mathcal{R}'$ into Eq.~\eqref{jones}, the corresponding Jones matrices are expressed as
\begin{align}
    \mathcal{J}(2k+1)=\left(\begin{array}{ccc}
1\\
 & 1\\
 &  & 0
\end{array}\right) \;
\mathcal{J}(2k)=\left(\begin{array}{ccc}
0\\
 & 0\\
 &  & 1
\end{array}\right),
\end{align}
where $k\in \mathbb{N}$. The $\mathcal{J}(2k+1)$ implies that odd harmonics are allowed in both $x$ and $y$ direction. 
Nevertheless, $\mathcal{J}(2k)$ serves as a conventional linear polarizer with an axis along $z$ \cite{fowles1989}, indicating the presence of linearly polarized even harmonics in the $z$ direction. 
Thereby, the $\mathcal{C}_{2(z)}$ along with the broken $\mathcal{M}_{z}$ ensures that the purely odd harmonics occur within the $x$-$y$ plane while the purely even ones appear in the $z$ direction. 
This explains the orientation-dependent HH spectra in Fig.~\ref{fig1}(d)-(e). When the LP pulse is polarized along the $x$ axis (a special case), we have $\mathcal{R}=\mathcal{C}_{2(z)}$ and $\mathcal{R}'=\mathcal{M}_{y}$. 
The Jones matrix for even harmonics [$\mathcal{J}(2k)$] remains unchanged, while that for odd harmonics [$\mathcal{J}(2k+1)$] becomes a linear polarizer with an axis along $x$. 
This is because $\mathcal{M}_{y}$ eliminates harmonic emission in the $y$ direction. 
As a result, odd harmonics manifest only in the $x$ direction while even harmonics still occur in the $z$ direction, as indicated by the gray curves shown in Fig.~\ref{fig1}(b)-(c).
For the LCP pulse in $x$-$y$ plane, the shared symmetries are
$\mathcal{R}=\mathcal{C}_{4(z)}$ and $\mathcal{R}'=E$. Then the Jones matrices read
\begin{align}
    \mathcal{J}(4k\pm1)=\left(\begin{array}{ccc}
\frac{1}{2} & \pm\frac{i}{2} & 0\\
\mp\frac{i}{2} & \frac{1}{2} & 0\\
0 & 0 & 0
\end{array}\right) \;
\mathcal{J}(4k)=\left(\begin{array}{ccc}
0\\
 & 0\\
 &  & 1
\end{array}\right),
\end{align}
where $\mathcal{J}(4k+1)$ is exactly the left circular polarizer, $\mathcal{J}(4k-1)$ represents the right circular polarizer, and $\mathcal{J}(4k)$ is the linear polarizer with $z$ axis \cite{fowles1989}. 
These conform to the selection rule observed from the red curves depicted in Fig.~\ref{fig1}(b)-(c).
Based on the analysis above, it is reasonable to infer that similar anomalous harmonics can be generated in other materials sharing the same space group as TaAs, for instance, the normal insulator Ba$_2$S$_3$ \cite{Yamaoka}.

\begin{figure}
	\centering
	\includegraphics[width=\columnwidth]{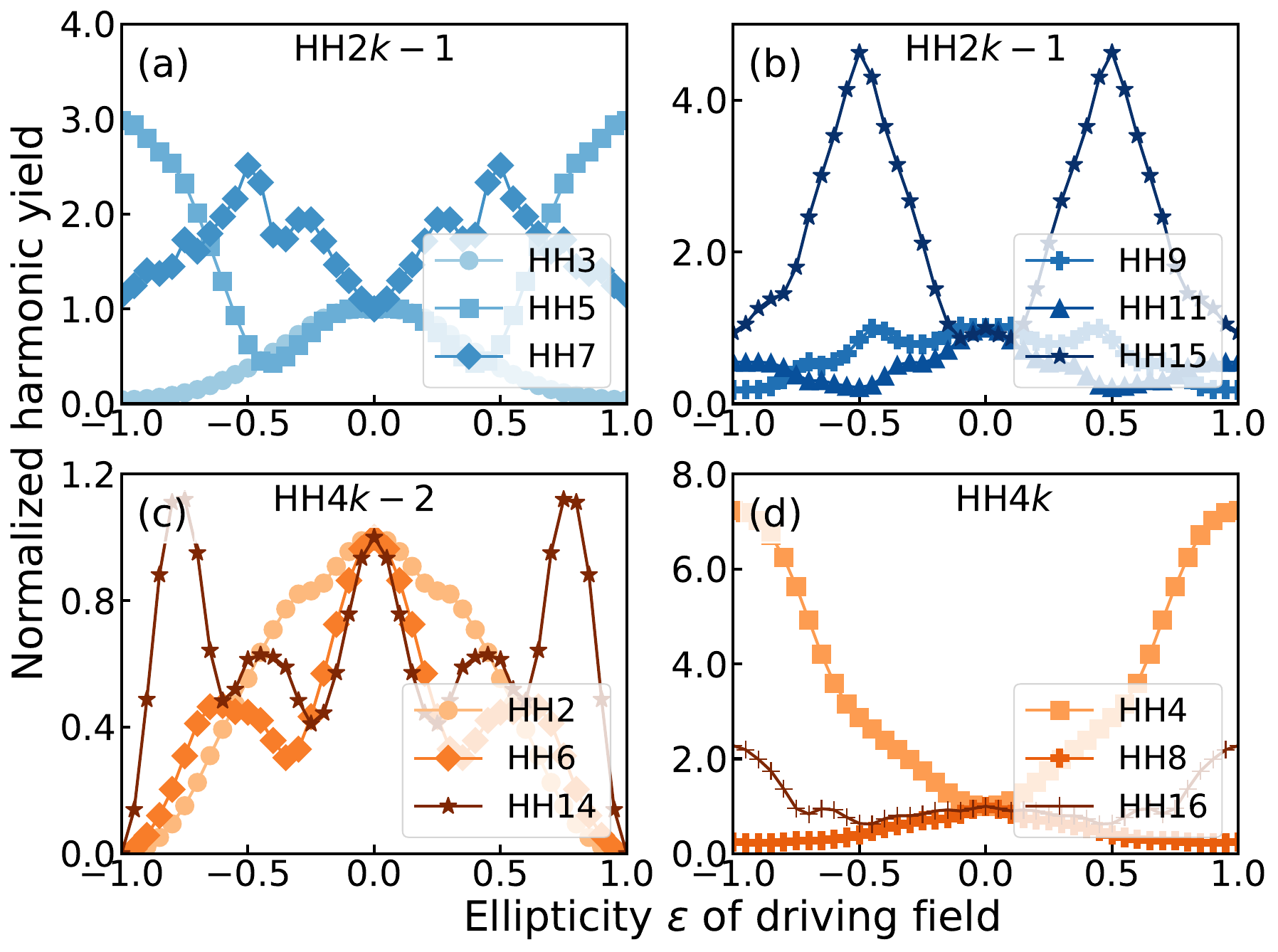}
	\caption{Normalized yields of HH$2k-1$ [(a) and (b)], HH$4k-2$ (c), and HH$4k$ (d) vs ellipticity of driving field. The laser parameters are the same as those in Fig.~\ref{fig1}.}
	\label{ellip}
\end{figure}

\textit{Impact of Weyl cones on HHG.}
Next, we shift our focus to the topological properties of TaAs.
Circular dichroism, as an approach of assessing chirality, is naturally considered for examining WPs with opposite topological charges.
In TaAs, there are $24$ WPs in total: $8$ WPs situated on the $k_z=0$ 
plane, referred to as $W_1$, and $16$ WPs located away from the $k_z=0$ plane, referred to as $W_2$ (see SM \cite{supp}). 
In each subset ($W_1$ or $W_2$), WPs are related to each other by $\mathcal{C}_{4(z)}$ and $\mathcal{T}$. 
Because of mirror symmetries ($\mathcal{M}_{x}$, $\mathcal{M}_{y}$, 
$\mathcal{M}_{xy}$, and $\mathcal{M}_{-xy}$), WPs from the same subset are at the same energy even with opposite chirality ($\chi$). 
The HHG produced by the interaction of a LCP laser with Weyl cone of $\chi=+1$ ($\chi=-1$) are thereby identical to those generated by 
the interaction of a right-handed circularly polarized (RCP) laser with  Weyl cone of $\chi=-1$ ($\chi=+1$). 
Considering the harmonic contributions across the entire BZ, the HH spectra excited by LCP and RCP pulses are the same. 
Consequently, the HHG CD in TaAs fails to probe the chirality of WPs, whatever the pulse polarization plane is chosen in the $x$-$y$ plane, $x$-$z$ plane, or $y$-$z$ plane.

In Fig.~\ref{ellip}, the yields of a few representative odd and even harmonics are plotted as a function of driver ellipticity.
Throughout the ellipticity scan (ranging from $-1$ to $1$), all harmonics exhibit a symmetric profile, 
with the major axis of the ellipse fixed along the $x$ axis. 
Specifically, most of the odd harmonics (excluding the $3$rd order) in Figs.~\ref{ellip}(a)-(b) show an anomalous ellipticity dependence (AED), where the yields of the harmonics are maximized for finite ellipticity. 
The even harmonics, however, present two distinctly different tendencies. 
In Fig.~\ref{ellip}(c), the ($4k-2$)th harmonics inevitably decay to zero under circular polarization, 
due to the selection rule constraints. 
In contrast, the majority of ($4k$)th (excluding the $8$th order) harmonics are enhanced with increasing ellipticity, illustrated in Fig.~\ref{ellip}(d). 
While the $8$th harmonic is suppressed under circular polarization, a complete decay is not observed.

\begin{figure}
	\centering
	\includegraphics[width=\columnwidth]{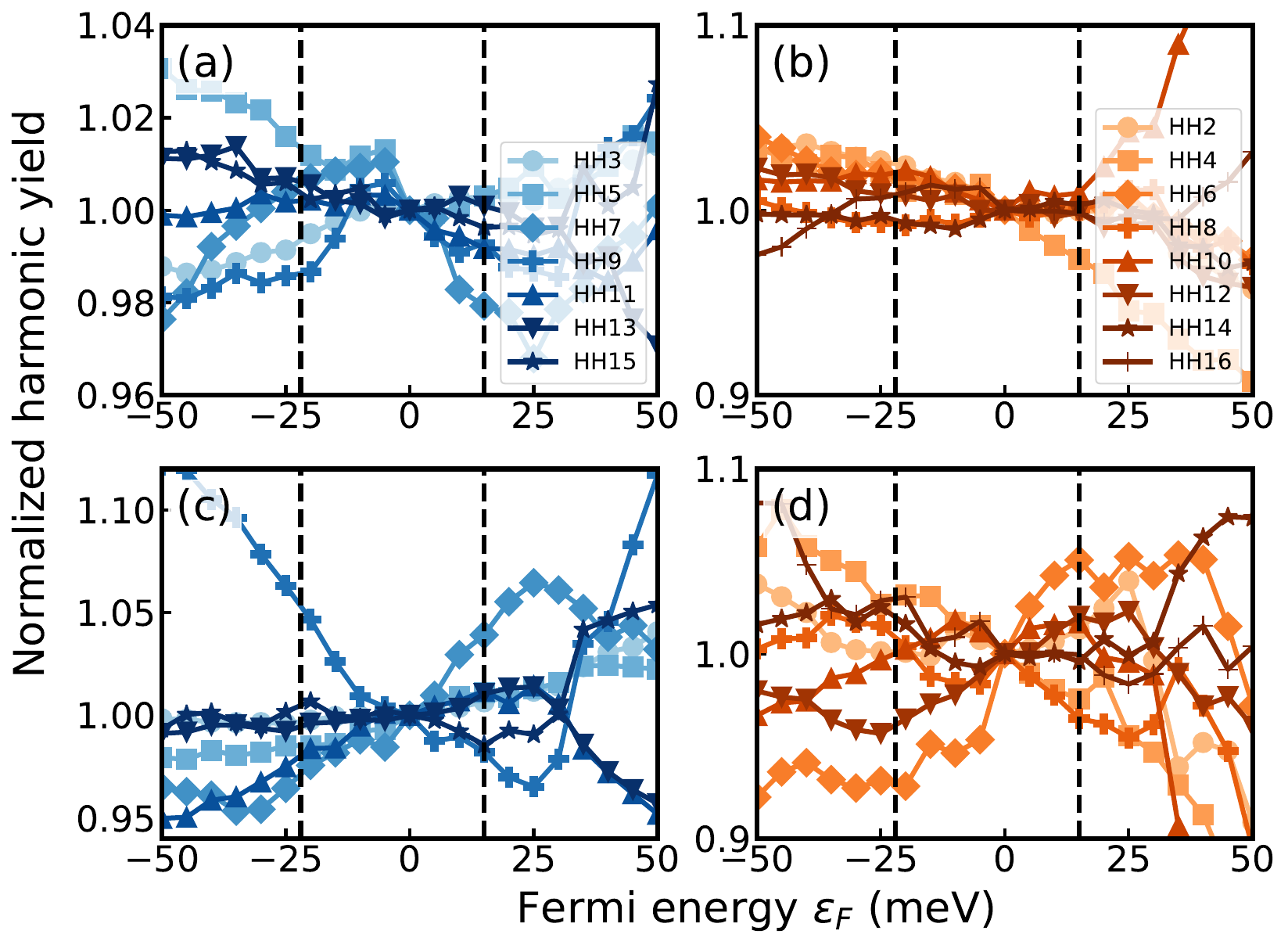}
	\caption{Normalized harmonic yields as a function of Fermi energy excited by linearly polarized fields with different wavelengths. The blue and orange curves represent odd and even harmonics, respectively. (a) and (b) The laser parameters are $\lambda=4200$ nm ($\hbar\omega=0.295$ eV), $I_0=5.04\times 10^{10}$ W/cm$^2$ ($E_0=0.0012$ a.u.), and $n_{\mathrm{cyc}}=20$. (c) and (d) The laser parameters are $\lambda=8400$ nm ($\hbar\omega=0.148$ eV), $I_0=1.26\times 10^{10}$ W/cm$^2$ ($E_0=0.0006$ a.u.), and $n_{\mathrm{cyc}}=20$. The black dashed lines indicate the energies of two nonequivalent sets of WPs ($W_1$ and $W_2$) in TaAs.}
	\label{hhgfermi}
\end{figure}

In graphene \cite{Yoshikawa736,PhysRevA.104.043525} and on the surface of 3D TIs \cite{PhysRevA.103.023101,Baykusheva2021nanoletter}, the AED of HHG is attributed to the presence of gapless Dirac cones.
However, in TaAs, we cannot establish such a causal relationship between the AED and the presence of Weyl cones.
i) HHG ellipticity dependence is highly sensitive to the laser parameters. Employing a pulse with lower intensity and shorter wavelength, we observed significant changes in the ellipticity dependence of HHG, with more harmonics exhibiting normal ellipticity dependence (see SM \cite{supp}). 
ii) In graphene, tuning the chemical potential suppresses the excitation near Dirac cones, leading to a transition from AED to normal ellipticity dependence \cite{PhysRevA.104.043525}. 
However, in TaAs, we employ the same approach to suppress the electron excitation near the Weyl cones. 
The ellipticity dependence of HHG remains anomalous (see SM \cite{supp}).
This shows that the Weyl cones do not dominate the HHG ellipticity dependence in TaAs.
iii) Furthermore, AED is extensively observed in solid-state systems, including insulators \cite{You2017,Tancogne-Dejean2017}, semiconductors \cite{PhysRevB.99.014304}, and semimetals \cite{Yoshikawa736,PhysRevA.104.043525,PhysRevA.97.063412}.
Therefore, it cannot serve as a fingerprint for identifying WSMs.

In addition to AED, the sensitivity of HHG intensity to excitations of Weyl fermions has been proposed as a possible way to differentiate Weyl cones with different topological charges \cite{PhysRevB.107.224308}.
Weyl semimetals feature the large tunability of the Fermi level \cite{PhysRevB.98.195446,Guo2023}. 
This provides a feasible experimental approach to block Weyl fermion excitations near the WPs by tuning the chemical potential.
In TaAs, the $W_1$ points are about $15$ meV above the Fermi energy ($\varepsilon_{F}=0$), while the $W_2$ points reside about $22$ meV
below the Fermi level.
Figure~\ref{hhgfermi} displays the normalized harmonic yields [$I(n\omega,\varepsilon_{F})/I(n\omega,\varepsilon_{F}=0)$] of both ordinary and anomalous harmonics as a function of Fermi energy, ranging from $-50$ to $50$ meV. 
We do not observe any obvious enhancement or reduction in the harmonic yields,
as the Fermi level scans across the energy of the WPs (black dashed lines) .
The yields of different orders fluctuate within a range of only $\pm 10\%$, 
for laser pulses with wavelengths of $4200$ nm [Figs.~\ref{hhgfermi}(a)-(b)] and $8400$ nm [Figs.~\ref{hhgfermi}(c)-(d)].
At $\varepsilon_{F}=50$ meV, there is no visible shift in the energy cutoff of the HH spectra (not show), although all WPs are Pauli blocked.
These imply that the intensity of HHG cannot be used to characterize the gapless excitations of electrons near the Weyl cones in TaAs.
\begin{figure}
	\centering
	\includegraphics[width=\columnwidth]{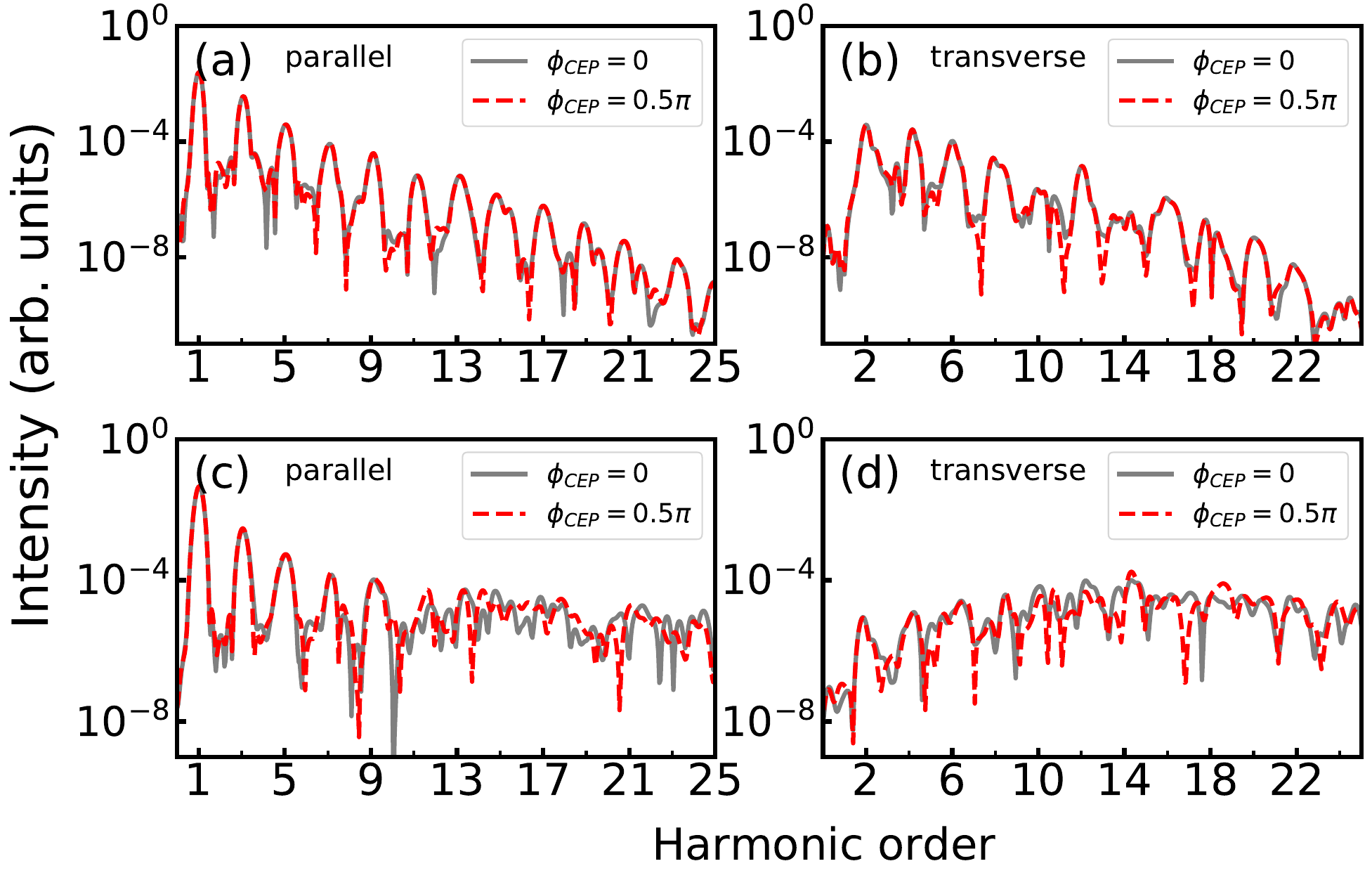}
	\caption{Parallel and transverse HH spectra excited by few-cycle ($n_{\mathrm{cyc}}=6$) pulses with different wavelengths. (a) and (b) Wavelength $\lambda=3200$ nm ($\hbar\omega=0.388$ eV), and $I_0=8.96\times 10^{10}$ W/cm$^2$ ($E_0=0.0016$ a.u.). (c) and (d) Wavelength $\lambda=8400$ nm ($\hbar\omega=0.148$ eV), and $I_0=4.24\times 10^{10}$ W/cm$^2$ ($E_0=0.0011$ a.u.). The gray dashed curves and red dashed curves correspond to $\phi_{\mathrm{CEP}} = 0$ and $\phi_{\mathrm{CEP}} = 0.5\pi$, respectively.}
	\label{cephhg}
\end{figure}

Moreover, the dependence of HHG on CEP is also insufficient for discerning the presence of Weyl points. This is because the HH spectra exhibits different sensivities to the variation of CEP,  when the TaAs driven by laser fields with different wavelengths. In Fig.~\ref{cephhg}(a), the HHG, excited by a LP laser with $\lambda=3200$ nm, is not sensitive to the CEP. However, a sensitive HHG CEP dependence occurs with increasing the wavelength to $\lambda=8400$ nm, as shown in Fig.~\ref{cephhg}(b).
This transition in CEP dependence extensively exsits in HHG from solids and is not exclusive to the HHG from WSMs. We attribute it to the change of the harmonic generation from multiphoton ionization-dominated to tunneling ionization-dominated \cite{PhysRevA.99.023406}.

\textit{Conclusion and outlook.}
In conclusion, we developed the Jones matrix formalism, providing a universal expression for HHG selection rules in crystalline solids. 
When applied to TaAs, we found that its anomalous even harmonics are induced by the $\mathcal{C}_{2(z)}$ symmetry, under the premise of breaking the $\mathcal{M}_{z}$ (or $\mathcal{I}$). 
It is noteworthy that non-topological materials with the same space group can also generate similar anomalous even harmonics. Hence, the anomalous HHG is not a reliable indicator of WSMs. 
Furthermore, we probed WPs in TaAs by using HHG CD, HHG ellipticity dependence, chemical potential-dependent HHG, and HHG CEP sensitivity. 
Our observations suggest that these common observables are more influenced by crystal symmetries and laser parameters rather than by the topologically protected Weyl cones. 
These indicate a similar challenge in detecting WSM topological features with HHG as with probing features of quantum spin/anomalous Hall insulators \cite{PhysRevX.13.031011}.

These findings for extracting topological features in TaAs can be extended into other semimetal materials, including the other type-I WSMs, type-II WSMs, and Dirac semimetals. In other words, it is impractical to investigate the electronic features of topological semimetals depending solely on harmonic signals induced by monochromatic fields at this stage. The multidimensional optical manipulation may also be a promising method of extracting topology of semimetals. For instance, two-color and multicolor laser fields might be able to maximize the contribution of  the electron-hole recombination events highly influenced by the Weyl cones on HHG. In chiral Weyl semimetals (lacking mirror symmetry), the chirality of WP is expected to be measured using HHG CD or HHG elliptical dichroism \cite{Juan2017,PhysRevB.108.L020305,sciadv.aba0509}. It is interesting to observe that even in nonchiral Weyl semimetals, such as TaAs, the chiral anomaly can be induced by applying appropriate external electromagnetic field \cite{Ma2017,PhysRevB.87.235306,PhysRevB.86.115133}, thus breaking mirror symmetries.

\textit{Acknowledgements.}
X.Z. thanks Ulrike Nitzsche for technical assistance.
This work is supported by the German Research Foundation (Deutsche
Forschungsgemeinschaft, DFG) via SFB1143 Project No. A5 and under Germany's Excellence Strategy
through W{\"u}rzburg‐Dresden Cluster of Excellence on Complexity and Topology in Quantum Matter ‐
\textit{ct.qmat} (EXC 2147, Project No. 390858490)

%


\end{document}


\title{{\mbt}}
\title{Supplementary Material: Impact of crystal symmetries and Weyl nodes on high-harmonic generation in Weyl semimetal \texorpdfstring{TaAs}{}}
\author{Xiao Zhang}
\affiliation{Institute for Theoretical Solid State Physics, Leibniz IFW Dresden, Helmholtzstrasse 20, 01069 Dresden, Germany}
\affiliation{Institute of Theoretical Physics and W{\"u}rzburg-Dresden  Cluster of Excellence {\it ct.qmat}, Technische Universit{\"a}t Dresden, 01062 Dresden, Germany} 
\author{Jeroen van den Brink}
\affiliation{Institute for Theoretical Solid State Physics, Leibniz IFW Dresden, Helmholtzstrasse 20, 01069 Dresden, Germany}
\affiliation{Institute of Theoretical Physics and W{\"u}rzburg-Dresden  Cluster of Excellence {\it ct.qmat}, Technische Universit{\"a}t Dresden, 01062 Dresden, Germany} 
\author{Jinbin Li}
\affiliation{Max-Planck-Institut f{\"u}r Physik komplexer Systeme, N{\"o}thnitzerstrasse 38, 01187 Dresden, Germany}
\affiliation{School of Nuclear Science and Technology, Lanzhou University, Lanzhou 730000, China}

\maketitle

In Supplementary~\ref{sec:methods}, we show the methods for Density functional theory (DFT), Wannier projection, and high-harmonic generation (HHG) calculation. 
Supplementary~\ref{sec:SOC} illustrates HHG in TaAs with and without spin-orbit coupling (SOC). 
Supplementary~\ref{sec:jonesmf} shows the derivation and application of Jones matrix formalism. 
In Supplementary~\ref{sec:invers}, we discuss the impact of inversion symmetry breaking on HHG by using a four-band minimal model of TaAs. 
Supplementary~\ref{sec:orenap} shows orientation dependence of HHG in TaAs. 
In supplementary~\ref{sec:CDich}, we calculate the circular dichrosim of HHG in TaAs with different polarization planes. 
Supplementary~\ref{sec:Ellip} exhibits ellipticity dependence of HHG, where the laser parameters and Fermi levels are different from those in the main text.

\section{\label{sec:methods}Methods}

\subsection{\label{sec:tb}DFT calculation, Wannier fit, and Weyl point searching}

\begin{figure}
	\centering
	\includegraphics[width=0.7\columnwidth]{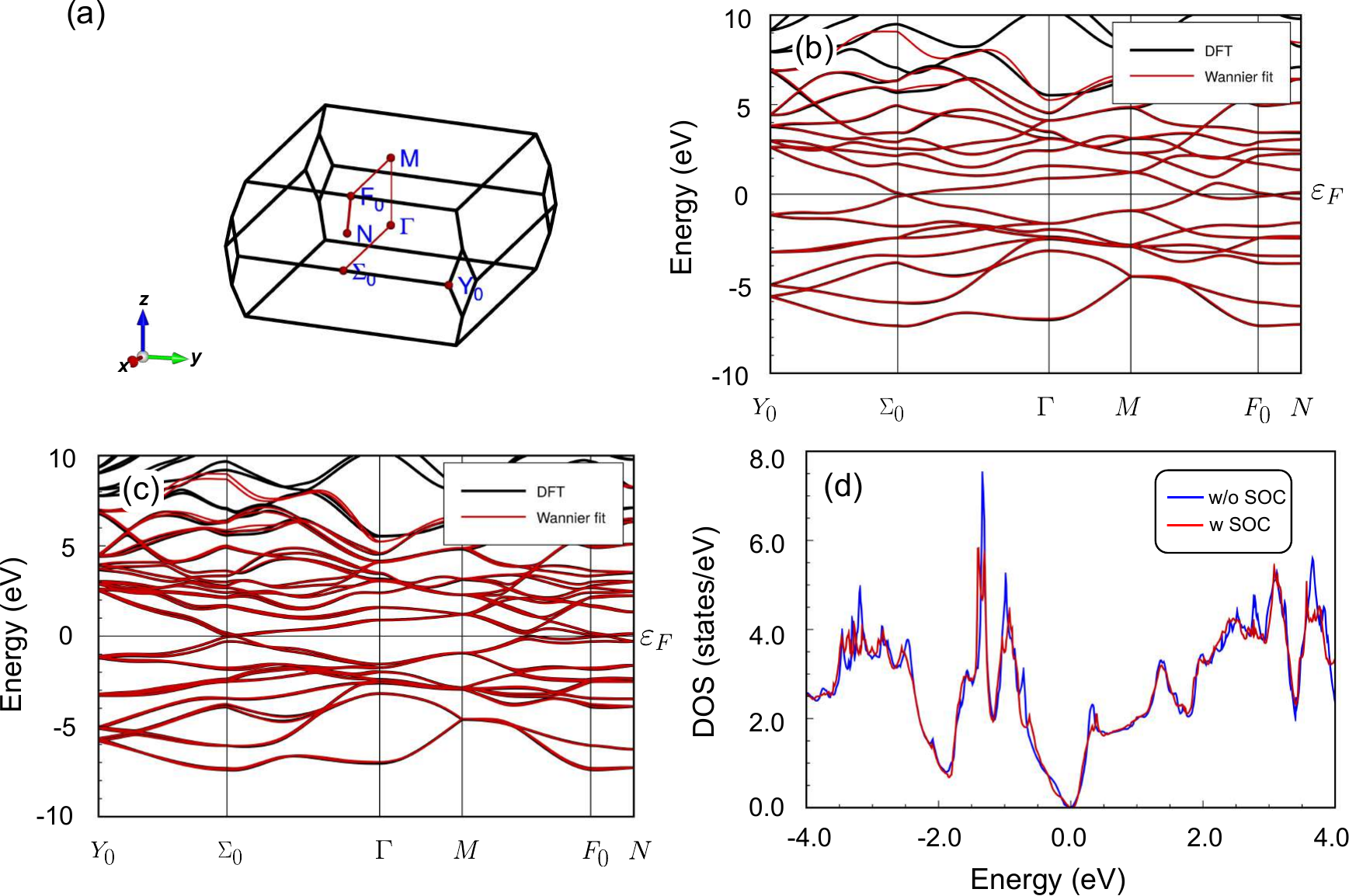}
	\caption{(a) Sketch of TaAs Brilloiun zone with high symmetry paths. (b) Band structure of TaAs without SOC by DFT (black curves) and Wannier fit (red curves). (c) Same as panel (b) but with SOC. (d) DOSs of TaAs without (blue curve) and with SOC (red curve).}
	\label{bandstr}
\end{figure}

The DFT calculations are carried out for TaAs using the Full-Potential Local-Orbital code (FPLO) code \cite{PhysRevB.59.1743}, version 21.61. \cite{fplo} The Perdew-Burke-Ernzerhof implementation of the generalized gradient approximation is employed in our calculations. For the numerical integration of DFT calculations, we use a  $k$ mesh with ($12\times 12 \times 12$) intervals in the Brillouin zone (BZ). The linear tetrahedron method is used for the numerical
integration in the BZ.
As applicable, the full relativistic correction is used, 
while the spin-orbit coupling effects are included via the 4-spinor formalism as implemented in the FPLO code.

\begin{table}
\setlength{\tabcolsep}{12pt}
\centering
\caption{The two nonequivalent Weyl points of TaAs. Positions, Chirality ($\chi$), and energies. Only the coordinates and Chirality of WPs in the first quadrant are given, the others can be obtained through symmetries.}
\label{table1}
\begin{tabular}{l r c r}
\hline\hline
 & $(k_x,k_y,k_z)$ in $\frac{2\pi}{a}$ & $\chi$ & $E$ (meV) \\
\hline
$W_1$ & $(0.048, 0.515, 0.000)$ & $+1$ & $15$ \\
$W_2$ & $(0.031, 0.236, 0.172)$ & $-1$ & $-22$ \\
\hline\hline
\end{tabular}
\label{tab:wps}
\end{table}

TaAs belongs to the space group $I4_{1}md$ (No.109) with lattice constants $a=b=3.4348$ {\AA} and $c=11.641$ {\AA}. 
The crystal structure consists of interpenetrating Ta and As sublattices, 
with two Ta atoms and two As atoms in each primitive unit cell. 
In Fig.~\ref{bandstr}(a), we show the BZ of TaAs with high-symmetry points and path. 
The energy bands (black curves), obtained by self consistent calculation without and with SOC, 
along the high-symmetry path are shown in Figs.~\ref{bandstr}(b) and (c), respectively. 
The corresponding density of states (DOSs) are plotted in Fig.~\ref{bandstr}(d).
 
To study the ultrafast dynamics in TaAs under a strong laser field,
the tight-binding Hamiltonian $(\hat{H}_{\mathrm{WF}}^{\boldsymbol{k}})$ and non-Abelian Berry connection $(\hat{\boldsymbol{D}}_{\mathrm{WF}}^{\boldsymbol{k}})$ are constructed by
projecting the Hamiltonian and position operator onto localized Wannier functions $\mid \boldsymbol{R}n\rangle$, respectively
\begin{align}
    (\hat{H}_{\mathrm{WF}}^{\boldsymbol{k}})_{nm}&=\sum_{\boldsymbol{R}}e^{i\boldsymbol{k}\cdot\boldsymbol{R}}\langle\boldsymbol{0}n\mid\hat{H}_{0}\mid\boldsymbol{R}m\rangle,\\
    (\hat{\boldsymbol{D}}_{\mathrm{WF}}^{\boldsymbol{k}})_{nm}&=\sum_{\boldsymbol{R}}e^{i\boldsymbol{k}\cdot\boldsymbol{R}}\langle\boldsymbol{0}n\mid\hat{\boldsymbol{r}}\mid\boldsymbol{R}m\rangle,
\end{align}
where $\boldsymbol{R}$ sums over all the lattice vectors.
The Wannier tight-binding model comprises the Ta-$5d$, $6s$ and As-$4p$ orbitals, 
with energy window lying between $-7.5$ eV and $+6.2$ eV
Therefore, the basis set for the Wannier projections consists of $18$ orbitals for the scalar relativistic case (without SOC). 
Correspondingly, the basis of Wannier functions involved $36$ orbitals 
when spin-orbit interactions were considered.
The energy bands (red curves), by Wannier fit, are depicted on the top of DFT bands, as illustrated in Figs.~\ref{bandstr}(b) and \ref{bandstr}(c).

\begin{figure}
	\centering
	\includegraphics[width=0.4\columnwidth]{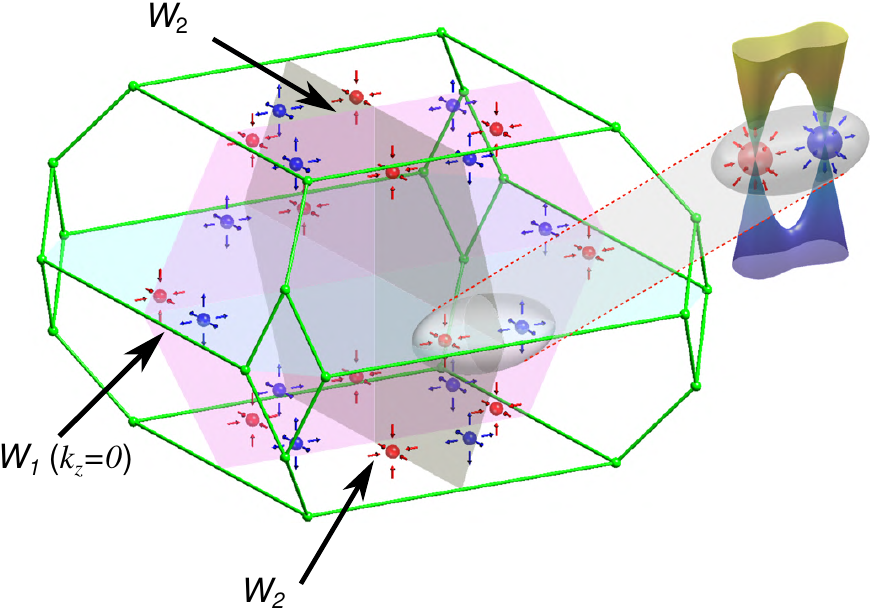}
	\caption{Weyl points in the Brillouin zone of TaAs with one elementary Weyl pair highlighted. Adapted from Ref.~\cite{PhysRevLett.123.246602}.}
	\label{WPs}
\end{figure}

The Weyl points (WPs) in BZ are searched and confirmed by computing the Chern
numbers as implemented in {\sc pyfplo} module of the FPLO code, and outlined
in Ref.~\cite{Klaus2016}. 
We find $24$ WPs in TaAs: $8$ WPs ($W_1$) situated on the $k_z=0$ 
plane, and $16$ WPs ($W_2$) located away from the $k_z=0$ plane, as shown in Fig.~\ref{WPs}.
The coordinates, chirality, and energies of WPs in TaAs are listed in Table~\ref{tab:wps}.

\subsection{\label{sec:method2}Time evolotion method for Wannier Hamiltonian}

To simulate the interaction between the laser field and the system, we solve the time-dependent Schr{\"o}dinger equation using the velocity gauge within the pseudo-Houston representation \cite{wannier_SBEs_yupeng}. This approach employs an accelerated Bloch-like basis formed by maximally localized Wannier functions
\begin{equation}
	i\frac{\partial}{\partial t}\left |m,\boldsymbol{k}_0,t\right>=\left[\hat{H}_{\mathrm{WF}}^{\boldsymbol{k}(t)}+\boldsymbol{F}(t)\cdot\hat{\boldsymbol{D}}_{\mathrm{WF}}^{\boldsymbol{k}(t)}\right]\left |m,\boldsymbol{k}_0,t\right>, \label{TDSE}
\end{equation}
where $m$ is the band index. The quasi-momentum is $\boldsymbol{k}(t)=\boldsymbol{k}_0+\boldsymbol{A}(t)$, with $\boldsymbol{k}_0$ being the initial quasi-momentum.  
The vector potential $\boldsymbol{A}(t)$ of laser field is expressed as
\begin{align}
\begin{split}
        \boldsymbol{A}(t)=\frac{E_{0}}{\omega_0}f(t)\left[\frac{1}{\sqrt{1+\epsilon^{2}}}\cos(\omega_0 t+\phi)\boldsymbol{e}_{x}
        +\frac{\epsilon}{\sqrt{1+\epsilon^{2}}}\sin(\omega_0 t+\phi)\boldsymbol{e}_{y}\right],
\end{split}
\label{eq:V_P}
\end{align}
where $E_0$ is the peak electric field, $f(t)=\sin^{2}{\left(\omega_0{t}/2n_{\mathrm{cyc}}\right)}$ is the envelope, $\omega_0$ is the fundamental frequency, $\varepsilon$ is the ellipticity, $\phi$ is the carrier envelope phase, and $n_{\mathrm{cyc}}$ is the total number of laser cycles. 
The corresponding electric field is $\boldsymbol{F}(t)=-d\boldsymbol{A}(t)/dt$ . 
We omit $\hat{\boldsymbol{D}}_{\mathrm{WF}}^{\boldsymbol{k}(t)}$ in this work, because it only has a minor influence on high-harmonic generation \cite{wannier_SBEs_yupeng}.

The wave functions of all occupied states are propagated independently by repetitively applying the time evolution operator until the end of the laser pulse
\begin{equation}
	\left |m,\boldsymbol{k}_0,t+\Delta t\right>=\hat{U}(t,t+\Delta t,\boldsymbol{k}_0)\left |m,\boldsymbol{k}_0,t\right>.\label{solution}
\end{equation}
Here, the time evolution operator is evaluated by 
\begin{equation}
	\hat{U}(t,t+\Delta t,\boldsymbol{k}_0)=\sum_{n}\left |n,\boldsymbol{k}(t)\right>e^{-i\lambda_{n}[\boldsymbol{k}(t)]\Delta t}\left <n,\boldsymbol{k}(t)\right|, \label{U_op}
\end{equation}
where $\lambda_{n}[\boldsymbol{k}(t)]$ and $\left |n,\boldsymbol{k}(t)\right>$ are the eigenvalues and eigenstates of the time-dependent Hamiltonian  $\hat{H}_{\mathrm{WF}}^{\boldsymbol{k}(t)}$. The time-dependent current reads
\begin{equation}
	\boldsymbol{J}(t)=\int_{\mathrm{BZ}}[d\boldsymbol{k}_{0}]\sum_{m}\left<m,\boldsymbol{k}_0,t\right|\hat{\boldsymbol{p}}\left[\boldsymbol{k}(t)\right]\left|m,\boldsymbol{k}_0,t\right>, \label{JtBZ}
\end{equation}
where $\int [d\boldsymbol{k}_{0}]$ denotes $\int d^{d}k_{0}/(2\pi)^{d}$, and $d$ indicates the dimension of system.
The momentum operator in the pseudo-Houston representation \cite{wannier_gauge} reads  
\begin{equation}
\hat{\boldsymbol{p}}=\nabla_{\boldsymbol{k}}\hat{H}_{\mathrm{WF}}^{\boldsymbol{k}(t)}+i[\hat{H}_{\mathrm{WF}}^{\boldsymbol{k}(t)},\hat{\boldsymbol{D}}_{\mathrm{WF}}^{\boldsymbol{k}(t)}],
\end{equation}
where $[\hat{H}_{\mathrm{WF}}^{\boldsymbol{k}(t)},\hat{\boldsymbol{D}}_{\mathrm{WF}}^{\boldsymbol{k}(t)}]$ represents the commutation of the Hamiltonian and position operator, and is omitted as mentioned above. 
The HHG spectrum is computed by 
\begin{equation}
I_{l}(\omega)=\left|\int dt e^{-i\omega t}W(t)\frac{d}{dt}J_{l}(t)\right|^2,
\end{equation}
where $l=x,y,z$ indicates the direction index, and $W(t)$ denotes the Blackman window function.
For the integral of Eq.~\eqref{JtBZ}, the BZ is sampled by $k$ grids from $20 \times 20 \times  20$ to $100 \times  100 \times  100$. Satisfactory convergence is achieved for a $k$ grid of size $50 \times  50 \times  50$ for HHG spectra.

\subsection{\label{sec:CD}Formulas of HHG ellipticity and HHG circular dichroism}
The vector potential [Eq.~\eqref{eq:V_P}] represents a right-handed circularly polarized (RCP) field when $\varepsilon=-1$, while it indicates a left-handed circularly polarized (LCP) field when $\varepsilon=+1$.
If we take into account the complex amplitude of vector potential,
the LCP field obeys $A_y/A_x=-i$ while the RCP field obeys $A_y/A_x=i$.

The ellipticity of harmonic emission reads
\begin{align}
    \varepsilon_{n}=\frac{\sqrt{I_{n}^{(+)}}-\sqrt{I_{n}^{(-)}}}{\sqrt{I_{n}^{(+)}}+\sqrt{I_{n}^{(-)}}},
\end{align}
where $I_{n}^{( \pm)}=\left(n \omega\right)^2\left|J_{x}\left(n \omega\right) \pm i J_{y}\left(n \omega\right)\right|^2$
represents the intensity of harmonic field projected
into two orthogonal ($\pm$) circularly polarized basis sets.
The circular dichroism is defined in terms of the circular components $I_{n, v}^{( \pm)}$ 
\begin{align}
    \mathrm{CD}_n=\frac{I_{n, \mathrm{RCP}}^{(+)}-I_{n, \mathrm{LCP}}^{(-)}}{I_{n, \mathrm{RCP}}^{(+)}+I_{n, \mathrm{LCP}}^{(-)}},\label{CDn}
\end{align}
where $v=\mathrm{RCP/LCP}$ in $I_{n, v}^{( \pm)}$ denotes the polarization state of driving fields.

\section{\label{sec:SOC}Effect of SOC}
\begin{figure}
	\centering
	\includegraphics[width=0.5\columnwidth]{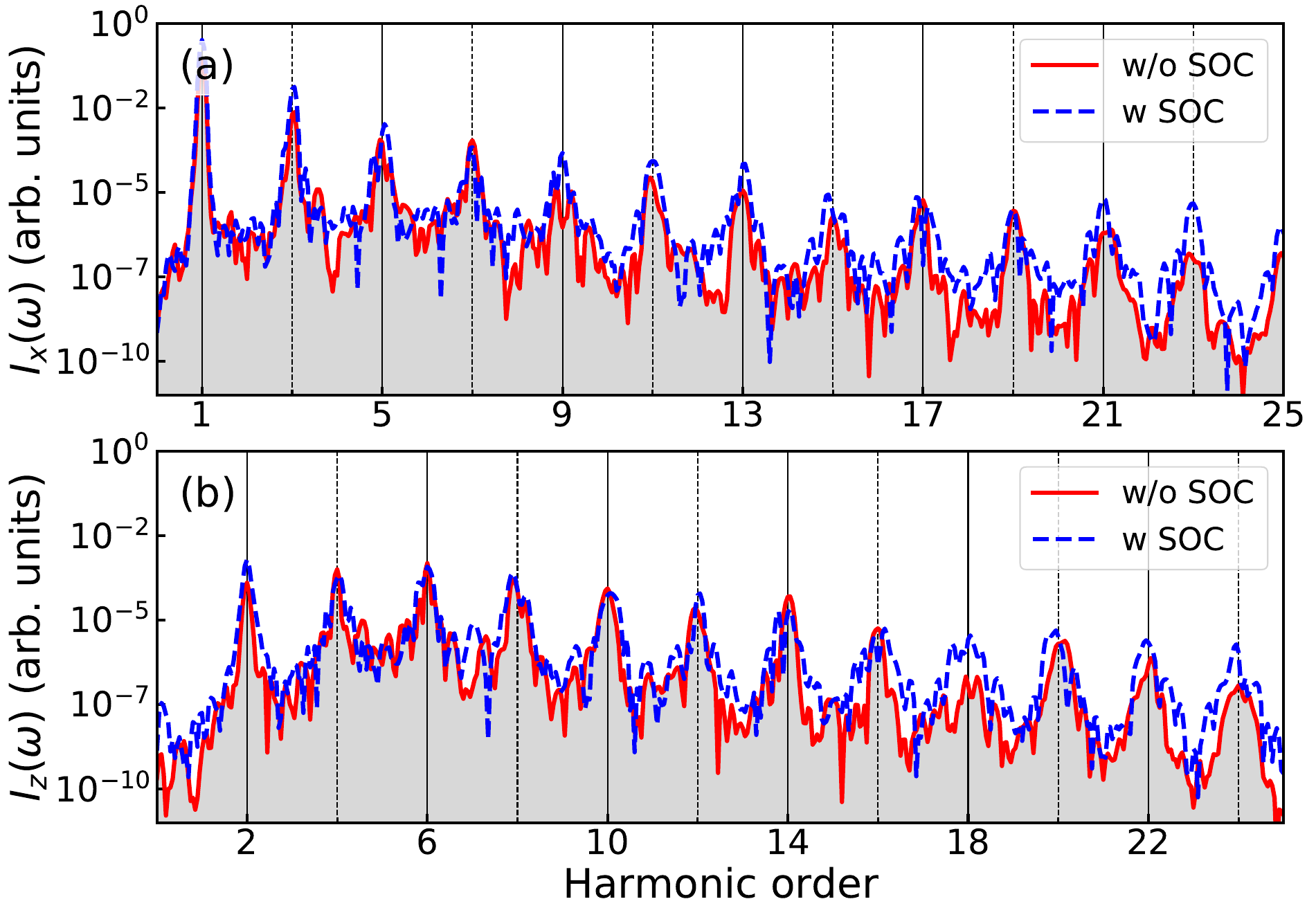}
	\caption{HHG spectra in TaAs without and with SOC driven by a linearly polarized (LP) pulse. (a) HHG in $x$ direction $I_x(\omega)$ and (b) HHG in $z$ direction $I_z(\omega)$. The red and blue curves represent harmonic spectra in TaAs with SOC and without SOC, respectively. The laser is polarized along the $x$ axis with intensity $I_{0} = 5.04 \times 10^{10}$ W/cm$^2$ ($E_{0} = 0.0012$ a.u.), wavelength $\lambda=4200$ nm ($\hbar\omega=0.295$ eV), and a duration of $20$ cycles.}
 \label{soc}
\end{figure}
Our DFT calculations consider two scenarios: the scalar relativistic approximation (without SOC) and the full relativistic correction (with SOC),
because the SOC is expected to play a crucial role in determining the properties of topological materials. In topological insulators, 
the absence of SOC can result in a trivial insulator or a Dirac semimetal phase (e.g., as in Bi$_2$Se$_3$, Bi$_2$Te$_3$, $1T'$-MoS$_2$, $1T'$-WTe$_2$, and so forth).
Activating SOC modifies the band gap and gives rise to a distinct  topological phase, significantly influencing the harmonic emissions.

However, the HHG in TaAs shows relatively minor effect from SOC, as observed in the comparison of the red and blue curves in Figs.~\ref{soc}(a)-(b).
This is because SOC leads to only slight alterations in the band structure and DOSs in TaAs, as illustrated in Figs.~\ref{bandstr}(b)-(d).
Note that the influence of SOC has been considered in all calculations in the main text unless otherwise stated.

\section{\label{sec:jonesmf}Jones matrix formalism}
\subsection{\label{sec:operator} Eigenvalue problems: the structural symmetry of the Hamiltonian}
Let $\mathcal{\hat{O}}$ be an operation of space group symmetry defined by a combination of translation (a fractional lattice vector) and orthogonal transformation $\mathcal{\hat{R}}$. The scalar $f(\boldsymbol{r})$ function is transformed by
\begin{align}
\mathcal{\hat{O}}f(\boldsymbol{r}) = f(\mathcal{R}^{-1}\boldsymbol{r}-\mathcal{R}^{-1}\frac{a_{j}}{M}),\label{eq:Tsf}
\end{align}
where $\mathcal{R}$ is the matrix representation of orthogonal transformation ($\mathcal{\hat{R}}$), $a_{j}$ indicates the $j$th lattice vector, and $M$ is an integer number.

The eigenvalue problem of periodic part of Bloch states is
\begin{equation}
   \hat{H}(\boldsymbol{k})\left|n,\boldsymbol{k}\right>=E_{n}^{\boldsymbol{k}}\left|n,\boldsymbol{k}\right>
   \label{eq:TIDSE}
\end{equation}
with the Hamiltonian 
\begin{equation}
     \hat{H}(\boldsymbol{k})=\frac{\hat{\boldsymbol{p}}^{2}}{2}+\frac{{\boldsymbol{k}}^{2}}{2}+\boldsymbol{k}\cdot\hat{\boldsymbol{p}}+\hat{V},
\label{eq:H0}
\end{equation}
where $\hat{\boldsymbol{p}}$ is the momentum operator, $\hat{V}$ is the potential operator,
and $E_{n}^{\boldsymbol{k}}$ is the eigenvalue of the state $\left|n,\boldsymbol{k}\right>$.
If the system is invariant under a space group transformation, the operator $\mathcal{\hat{O}}$ commutes
with the Hamiltonian:
\begin{align}
\begin{split}
\mathcal{\hat{O}}\hat{H}(\boldsymbol{k})\mathcal{\hat{O}}^{-1}&=\frac{\hat{\boldsymbol{p}}^{2}}{2}+\frac{{\boldsymbol{k}}^{2}}{2}+\boldsymbol{k}\cdot\mathcal{\hat{O}}\hat{\boldsymbol{p}}\mathcal{\hat{O}}^{-1}+\hat{V}\\
&=\frac{\hat{\boldsymbol{p}}^{2}}{2}+\frac{\boldsymbol{k}^{2}}{2}+\boldsymbol{k}\cdot\mathcal{\hat{R}}\hat{\boldsymbol{p}}\mathcal{\hat{R}}^{-1}+\hat{V}\\
&=\frac{\hat{\boldsymbol{p}}^{2}}{2}+\frac{(\mathcal{R}\boldsymbol{k})^{2}}{2}+\mathcal{R}^{-1}(\mathcal{R}\boldsymbol{k})\cdot\mathcal{R}^{-1}\hat{\boldsymbol{p}}+\hat{V}\\
&=\hat{H}(\mathcal{R}\boldsymbol{k}).\label{eq:H}
\end{split}
\end{align}
Here we have considered
\begin{align}
    \begin{split}
        \mathcal{\hat{R}}\hat{\boldsymbol{p}}\mathcal{\hat{R}}^{-1}f(\boldsymbol{r})&=-\mathcal{\hat{R}}i\nabla_{\boldsymbol{r}}f(\mathcal{R}\boldsymbol{r})\\ &=-\mathcal{\hat{R}}\mathcal{R}^{-1}i\nabla_{\mathcal{R}\boldsymbol{r}}f(\mathcal{R}\boldsymbol{r})\\
        &=-\mathcal{R}^{-1}i\left[\mathcal{\hat{R}}f_x(\mathcal{R}\boldsymbol{r}),\mathcal{\hat{R}}f_y(\mathcal{R}\boldsymbol{r}),\mathcal{\hat{R}}f_z(\mathcal{R}\boldsymbol{r})\right]^T\\
&=\mathcal{R}^{-1}\hat{\boldsymbol{p}}f(\boldsymbol{r})
    \end{split}
    \label{eq:nabla}
\end{align}
and the dot product of any two vectors is invariant under the orthogonal transformation.
Applying  $\mathcal{\hat{O}}$ to the two side of Eq. \eqref{eq:TIDSE}, we find 
\begin{align}
\begin{split}
\mathcal{\hat{O}}\hat{H}(\boldsymbol{k})\left|n,\boldsymbol{k}\right>&=\mathcal{\hat{O}}\hat{H}(\boldsymbol{k})\mathcal{\hat{O}}^{-1}\mathcal{\hat{O}}\left|n,\boldsymbol{k}\right>\\
&=\hat{H}(\mathcal{R}\boldsymbol{k})\mathcal{\hat{O}}\left|n,\boldsymbol{k}\right>\\
&=E_{n}^{\boldsymbol{k}}\mathcal{\hat{O}}\left|n,\boldsymbol{k}\right>.\label{eq:TIDSE1}
\end{split}
\end{align}
Due to $\mathcal{\hat{O}}\left|n,\boldsymbol{k}\right>$ is the eigenstate of $\hat{H}(\mathcal{R}\boldsymbol{k})$ with the eigenvalue $E_{n}^{\boldsymbol{k}}$,
the eigenvalue and eigenstate satisfy
\begin{align}
\begin{split}
   &E_{n}^{\boldsymbol{k}}=E_{n}^{\mathcal{R}\boldsymbol{k}},\\
    &\mathcal{\hat{O}}\left|n,\boldsymbol{k}\right>=\left|n,\mathcal{R}\boldsymbol{k}\right>e^{i\phi_n^{\mathcal{R}\boldsymbol{k}}},\label{eq:state}
\end{split}
\end{align}
where $\phi_n^{\mathcal{R}\boldsymbol{k}}$ is a gauge dependent phase factor.

\subsection{\label{sec:Sel.} Jones matrix for HHG under structural symmetry}
In this section, our goal is to derive the Jones matrix for HHG (from crystalline solids) in the following form
\begin{align}
    \mathcal{J}(n_{H})=\sum_{l'=1}^{N'}(\mathcal{R}')^{l'}\sum_{l=1}^{N}\exp(-i2\pi n_{H}\frac{l}{N})(\mathcal{R})^{l}, \label{jms}
\end{align}
where $n_{H}$ is the harmonic order, $\mathcal{R}$ and $\mathcal{R}'$ are the shared orthogonal transformations of external field and crystal structure, $N$ and $N'$ are respectively the orders of symmetry elements $\mathcal{R}$ and $\mathcal{R}'$ [$\mathcal{R}^{N}=E$, $(\mathcal{R}')^{N}=E$], and $E$ is the identity element. In the following derivation, we explain the difference of $\mathcal{R}$ and $\mathcal{R}'$.

We start our derivation here. Using Eq.~\eqref{JtBZ}, the amplitude of the $n_{H}$th harmonic is expressed as 
\begin{equation}
    \boldsymbol{J}(n_{H}\omega_{0})=\sum_{E_{n}^{\boldsymbol{k}}\leqslant\varepsilon_{F}}\int_{\mathrm{BZ}}[d\boldsymbol{k}_{0}]\int_{-\infty}^{+\infty}dt\left<n,\boldsymbol{k}_{0},t_{0}\right|\hat{U}^{\dagger}(\boldsymbol{k}_{0},t,t_{0})[\hat{\boldsymbol{p}}+\boldsymbol{k}_{0}+\boldsymbol{A}(t)]\hat{U}(\boldsymbol{k}_{0},t,t_{0})\left|n,\boldsymbol{k}_{0},t_{0}\right>e^{-in_{H}\omega_{0}t},\label{eq:J_w}
\end{equation}
where $\varepsilon_{F}$ is the Fermi energy, time evolution operator reads
\begin{align}
\hat{U}(\boldsymbol{k}_0,t,t_0)&= \hat{\mathcal{T}} e^{-i\int_{t_0}^{t}d\tau \hat{H}[\boldsymbol{k}_0+\boldsymbol{A}(\tau)]},
\label{eq:hatU}
\end{align}
and $\hat{\mathcal{T}}$ is the time ordering operator.
The generation of a specific harmonic is determined by the shared symmetries of the laser field and the lattice. 
These shared symmetries are classified into two categories in the following discussion: i) where the laser field under the symmetry operation can be transformed into the operation of temporal translation, and ii) where the laser field remains unchanged under the symmetry operation.

First, we discuss the case i).
We neglect $\boldsymbol{k}_0+\boldsymbol{A}(t)$ in Eq.~\eqref{eq:J_w} because it only contributes to the first order harmonic. 
The space group operation $\mathcal{\hat{O}}$ [defined by Eq. \eqref{eq:Tsf}] is equivalent to the orthogonal transformation $\mathcal{R}$ when applying to an external field. The external field is translated by $lT_{0}/N$ in time
\begin{equation}
    \mathcal{R}^{l}\boldsymbol{A}(t)=\boldsymbol{A}(t+lT_{0}/N),\label{eq:f_sym}
\end{equation}
when it is operated by the $l$th order of $\mathcal{R}$, where $T_{0}=2\pi/\omega_{0}$ is the period of laser field.
For the time-dependent Hamiltonian and time evolution operator [using Eq.~\eqref{eq:H}], we have
\begin{equation}
   \hat{H}[\boldsymbol{k}_0+A(\tau+lT_{0}/N)]=\hat{H}[\mathcal{R}^l\mathcal{R}^{-1}\boldsymbol{k}_0+\mathcal{R}^lA(\tau)]=\mathcal{\hat{O}}^{l}\hat{H}(\mathcal{R}^{-l}\boldsymbol{k}_0,\tau)\mathcal{\hat{O}}^{-l},
\end{equation}
and 
\begin{align}
\hat{U}(\boldsymbol{k}_0,t+lT_{0}/N,t_0+lT_{0}/N)=\mathcal{\hat{O}}^{l}\hat{U}(\mathcal{R}^{-l}\boldsymbol{k}_0,t,t_0)\mathcal{\hat{O}}^{-l}.
\label{eq:u_symm}
\end{align}
Equation~\eqref{eq:u_symm} allows us to reduce the integral over the entire time domain to $[0,T_0/N]$. The whole integral can be transferred into the summation of small intervals by $\int_{-\infty}^{\infty}dt \xrightarrow{} \sum_{m}\sum_{l=1}^{N}\int_{mT_0+(l-1)T_{0}/N}^{mT_0+lT_{0}/N}dt$. Thus, equation~\eqref{eq:J_w} can be rewritten by
\begin{align}
\begin{split}
    \boldsymbol{J}(n_{H}\omega_{0})=\sum_{E_{n}^{\boldsymbol{k}}\leqslant\varepsilon_{F}}\sum_{m}\sum_{l=1}^{N}\int_{\mathrm{BZ}}&[d\boldsymbol{k}_0]\int_{mT_0+(l-1)T_0/N}^{mT_0+lT_0/N}dt\\
    &\left<n,\boldsymbol{k}_0,t_0\right|\hat{U}^{\dagger}(\boldsymbol{k}_0,t,t_0)\hat{\boldsymbol{p}}\hat{U}(\boldsymbol{k}_0,t,t_0)\left|n,\boldsymbol{k}_0,t_0\right>e^{-in_{H}\omega_0t}.\label{eq:n_H_1}
\end{split}
\end{align}
Let $t_0=mT_0+(l-1)T_0/N$, the $n_{H}$th harmonic field becomes
\begin{align}
\begin{split}
      \boldsymbol{J}(n_{H}\omega_{0})=&\sum_{E_{n}^{\boldsymbol{k}}\leqslant\varepsilon_{F}}\sum_{m}\sum_{l=1}^{N}\int_{\mathrm{BZ}}[d\boldsymbol{k}_0]\int_{0}^{T_0/N}dte^{-in_{H}\omega_0t-in_{H}(l-1)2\pi/N}\\
&\left<n,\boldsymbol{k}_0,mT_0+(l-1)T_0/N\right|\hat{U}^{\dagger}(\boldsymbol{k}_0,t+(l-1)T_0/N,(l-1)T_0/N)\\
&\hat{\boldsymbol{p}}\hat{U}(\boldsymbol{k}_0,t+(l-1)T_0/N,(l-1)T_0/N)\left|n,\boldsymbol{k}_0,mT_0+(l-1)T_0/N\right>,
\label{eq:J_w_t}
\end{split}
\end{align}
by considering $\int_{mT_0+(l-1)T_{0}/N}^{mT_0+lT_{0}/N}f(t)dt=\int_{0}^{T_{0}/N}f[t+mT_0+(l-1)T_{0}/N]dt$ and $\hat{U}(\boldsymbol{k}_0,t+mT_{0},t_0+mT_{0})=\hat{U}(\boldsymbol{k}_0,t,t_0)$. 
Substituting Eqs.~\eqref{eq:f_sym} and \eqref{eq:u_symm} into Eq.~\eqref{eq:J_w_t}, we obtain
\begin{align}
\begin{split}
\boldsymbol{J}(n_{H}\omega_{0})=&\sum_{E_{n}^{\boldsymbol{k}}\leqslant\varepsilon_{F}}\sum_{m}\sum_{l=1}^{N}\int_{\mathrm{BZ}}[d\boldsymbol{k}_0]\int_{0}^{T_0/N}dte^{-in_{H}\omega_0t-in_{H}(l-1)2\pi/N}\\
&\left<n,\boldsymbol{k}_0,mT_0+(l-1)T_0/N\right|\mathcal{\hat{O}}^{(l-1)}\hat{U}^{\dagger}(\mathcal{R}^{-(l-1)}\boldsymbol{k}_0,t,0)\mathcal{\hat{O}}^{-(l-1)}\\
&\hat{\boldsymbol{p}}\mathcal{\hat{O}}^{(l-1)}\hat{U}(\mathcal{R}^{-(l-1)}\boldsymbol{k}_0,t,0)\mathcal{\hat{O}}^{-(l-1)}\left|n,\boldsymbol{k}_0,mT_0+(l-1)T_0/N\right>.
\label{eq:J_w_tt}
\end{split}
\end{align}
Assuming the low excitation of valence electrons, we can roughly think that the wave function propagates adiabatically on the valence bands
\begin{align}
\left|n,\boldsymbol{k}_0,mT_0+(l-1)T_0/N\right>\approx\left|n,\boldsymbol{k}_0+\boldsymbol{A}[mT_0+(l-1)T_0/N]\right>e^{i\phi_{n}^{\boldsymbol{k}_0}[mT_0+(l-1)T_0/N]}.\label{eq:wave}
\end{align}
Substituting Eqs.\eqref{eq:nabla},  \eqref{eq:state}, \eqref{eq:f_sym} and \eqref{eq:wave} into Eq.~\eqref{eq:J_w_tt}, we have
\begin{align}
\begin{split}
\boldsymbol{J}(n_{H}\omega_{0})=&\sum_{l=1}^{N}\mathcal{R}^{(l-1)}e^{-in_{H}(l-1)2\pi/N}\sum_{n,m}\int_{\mathrm{BZ}}[d\boldsymbol{k}'_0]\int_{0}^{T_0/N}dte^{-in_{H}\omega_0t}\\
&\left<n,\boldsymbol{k}'_0+\boldsymbol{A}(0)\right|
\hat{U}^{\dagger}(\boldsymbol{k}'_0,t,0)\hat{\boldsymbol{p}}\hat{U}(\boldsymbol{k}'_0,t,0)
\left|n,\boldsymbol{k}'_0+\boldsymbol{A}(0)\right>,\label{eq:selec0}
\end{split}
\end{align}
where  $\mathcal{R}^{-(l-1)}\boldsymbol{k}_0$ is replaced by $\boldsymbol{k}'_0$. 

Then we consider case ii): the laser field is invariant under the operator $\hat{\mathcal{O}}'$. The vector potential of the laser field satisfies
\begin{equation}
    \boldsymbol{A}(t)=\mathcal{R}'\boldsymbol{A}(t),
\end{equation}
and the potential operator satisfies
\begin{equation}
    \hat{V}=\hat{\mathcal{O}'}^{l}\hat{V}\hat{\mathcal{O}'}^{-l}.
\end{equation}
We give some specific examples of this case in the next section. Using Eqs.~\eqref{eq:H}, we immediately have 
\begin{align}
\hat{U}(\mathcal{R'}^{l}\boldsymbol{k}_0,t,t_0)=\mathcal{\hat{O'}}^{l}\hat{U}(\boldsymbol{k}_0,t,t_0)\mathcal{\hat{O'}}^{-l},
\end{align}
and
\begin{align}
\begin{split}
&\left<n,\mathcal{R'}^{l}\boldsymbol{k}'_0+\boldsymbol{A}(0)\right|
\hat{U}^{\dagger}(\mathcal{R'}^{l}\boldsymbol{k}'_0,t,0)\hat{\boldsymbol{p}}\hat{U}(\mathcal{R'}^{l}\boldsymbol{k}'_0,t,0)
\left|n,\mathcal{R'}^{l}\boldsymbol{k}'_0+\boldsymbol{A}(0)\right>\\
&=\left<n,\boldsymbol{k}'_0+\boldsymbol{A}(0)\right|
\hat{U}^{\dagger}(\boldsymbol{k}'_0,t,0)\mathcal{\hat{O'}}^{-l}\hat{\boldsymbol{p}}\mathcal{\hat{O'}}^{l}\hat{U}(\boldsymbol{k}'_0,t,0)
\left|n,\boldsymbol{k}'_0+\boldsymbol{A}(0)\right>\\
&=\mathcal{R'}^{l}\left<n,\boldsymbol{k}'_0+\boldsymbol{A}(0)\right|
\hat{U}^{\dagger}(\boldsymbol{k}'_0,t,0)\hat{\boldsymbol{p}}\hat{U}(\boldsymbol{k}'_0,t,0)
\left|n,\boldsymbol{k}'_0+\boldsymbol{A}(0)\right>. \label{eq:tevl}
\end{split}
\end{align}
Plugging Eq.~\eqref{eq:tevl} into \eqref{eq:selec0}, the integral over the entire first BZ converts to the integral over an irreducible Brillouin zone ($\mathrm{BZ}'$). Therefore the $n_{H}$th harmonic field is expressed by
\begin{align}
\begin{split}
\boldsymbol{J}(n_{H}\omega_{0})=&\mathcal{J}(n_{H})\boldsymbol{J}^{\mathrm{ir}}(n_{H}\omega_{0}),\label{eq:jones}
\end{split}
\end{align}
where
\begin{align}
\mathcal{J}(n_{H})=\sum_{l'=1}^{N'}\mathcal{R'}^{l'}\sum_{l=1}^{N}\mathcal{R}^{(l-1)}e^{-in_{H}(l-1)2\pi/N}
\label{eq:selec1}
\end{align}
and 
\begin{align}
    \boldsymbol{J}^{\mathrm{ir}}(n_{H}\omega_{0})=\sum_{n,m}\int_{\mathrm{BZ}'}[d\boldsymbol{k}'_0]\int_{0}^{T_0/N}dte^{-in_{H}\omega_0t}\left<n,\boldsymbol{k}'_0+\boldsymbol{A}(0)\right|
\hat{U}^{\dagger}(\boldsymbol{k}'_0,t,0)\hat{\boldsymbol{p}}\hat{U}(\boldsymbol{k}'_0,t,0)
\left|n,\boldsymbol{k}'_0+\boldsymbol{A}(0)\right>. \label{eq:irrcur}
\end{align}
We surprisingly find that $\mathcal{J}(n_{H})$ which induced by the structural symmetries is just the Jones matrix extensively used in the classic optics \cite{fowles1989}. It determines the allowed orders and the polarization states of HHG. Moreover, 
$\boldsymbol{J}^{\mathrm{ir}}(n_{H}\omega_{0})$ is the irreducible part of $n_{H}$th harmonic field and can be treated as a Jones vector.

One may note that shifting the summation range with respect to $l$ does not change the summation result in Eq.~\eqref{eq:selec1}. Thus we can convert Eq.~\eqref{eq:selec1} to Eq.~\eqref{jms}.

\subsection{\label{sec:EJMa}Examples of applying Jones matrix to TaAs HHG} 
In this section, we use Jones matrix [Eq.~\eqref{jms}] to provide a detailed explanation of the selection rule for TaAs HHG shown in main text. Then we give an extra example of HHG excited by LP laser along $z$ direction.

\subsubsection{LP driver in x-y plane}
\begin{figure}
	\centering
	\includegraphics[width=0.75\columnwidth]{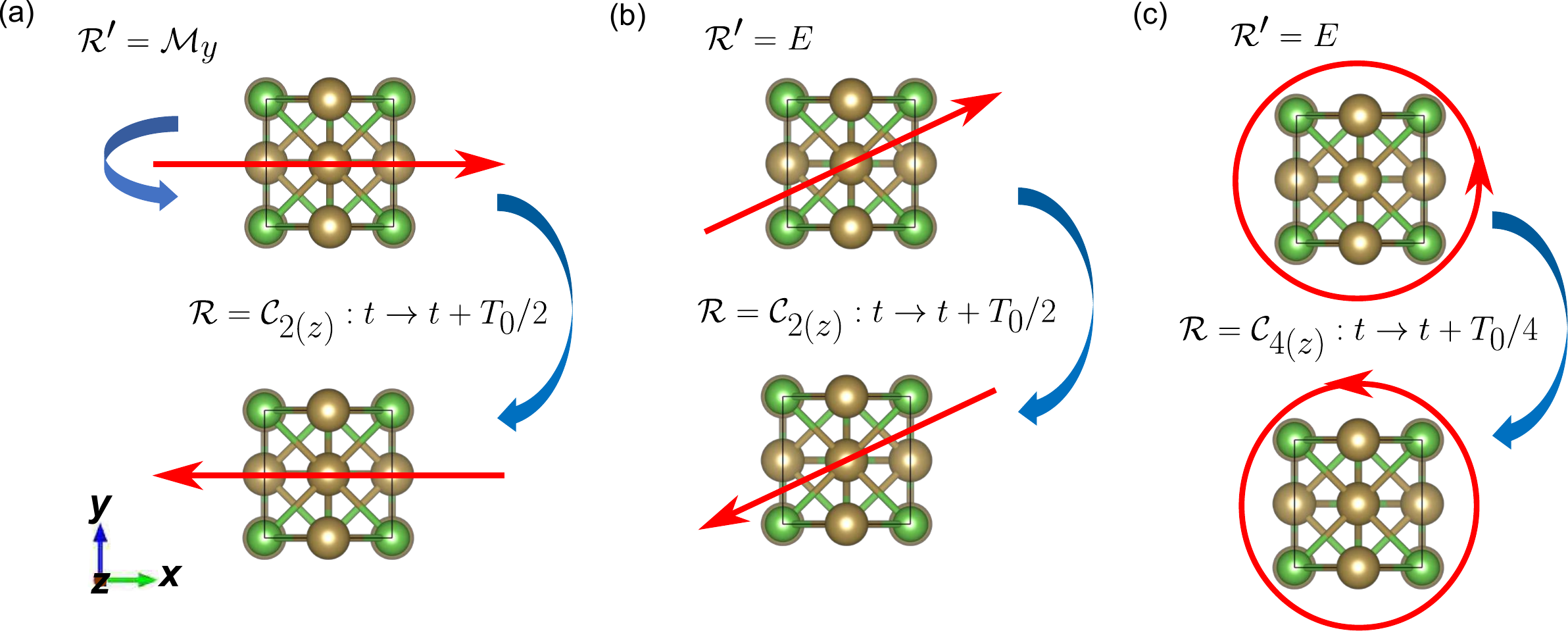}
	\caption{Sketch illustrating the interaction of laser field and TaAs lattice under operation of crystal symmetry.}
 \label{laserjones}
\end{figure}

Firstly, we focus on the case where the laser is linearly polarized along the $x$ direction as shown in Fig.~\ref{laserjones}(a). 
We find that the shared symmetries are $\mathcal{R}=\mathcal{C}_{2(z)}$ and $\mathcal{R}'=\mathcal{M}_{y}$. The vector potential $\boldsymbol{A}(t)$ is shifted by $T_0/2$ in time by the operation of $\mathcal{C}_{2(z)}$ but is invariant under the operation of $\mathcal{M}_{y}$.
The matrix representation of shared symmetries are
\begin{align}
\mathcal{R}=\mathcal{C}_{2(z)}=\left(\begin{array}{ccc}
-1\\
 & -1\\
 &  & 1
\end{array}\right)
\text{ and }
     \mathcal{R}'=\mathcal{M}_{y}=\left(\begin{array}{ccc}
1\\
 & -1\\
 &  & 1
\end{array}\right),\label{Rmy}
\end{align}
where $\mathcal{C}_{2(z)}(x,y,z)^{T}=(-x,-y,z)^{T}$ and $\mathcal{M}_{y}(x,y,z)^{T}=(x,-y,z)^{T}$. 
Plugging Eq.~\eqref{Rmy} into Eq.~\eqref{jms}, the corresponding Jones matrix is
\begin{align}
    \mathcal{J}(n_{H})=2\left(\begin{array}{ccc}
1-e^{-i\pi n_{H}}\\
 & 0\\
 &  & 1+e^{-i\pi n_{H}}
\end{array}\right). \label{c2jon}
\end{align}
Therefore, both odd ($n_H=2k+1$) and even ($n_H=2k$) harmonics are allowed according to Eq.~\eqref{c2jon}, where $k\in \mathbb{N}$. The Jones matrix for odd harmonics is 
\begin{align}
    \mathcal{J}(n_{H}=2k+1)=4\left(\begin{array}{ccc}
1\\
 & 0\\
 &  & 0
\end{array}\right),
\end{align}
while the Jones matrix for even harmonics is
\begin{align}
    \mathcal{J}(n_{H}=2k)=4\left(\begin{array}{ccc}
0\\
 & 0\\
 &  & 1
\end{array}\right).
\end{align}
From the perspective of classic optics, $\mathcal{J}(2k+1)$ is just a linear polarizer with an axis along $x$, while $\mathcal{J}(2k)$ serves as a linear polarizer with an axis along $z$. Applying Jones matrices to the corresponding Jones vectors, we find
\begin{align}
    \begin{split}
        \boldsymbol{J}[(2k+1)\omega_{0}]	&=\mathcal{J}(2k+1)\boldsymbol{J}^{ir}\\
	&=\left(\begin{array}{ccc}
1\\
 & 0\\
 &  & 0
\end{array}\right)\left(\begin{array}{c}
J_{x}^{ir}\\
J_{y}^{ir}\\
J_{z}^{ir}
\end{array}\right)
=J_{x}^{ir}[(2k+1)\omega_{0}]\boldsymbol{e}_{x},
    \end{split}\label{J2kp1}
\end{align}
and
\begin{align}
    \begin{split}
        \boldsymbol{J}[(2k)\omega_{0}]	&=\mathcal{J}(2k)\boldsymbol{J}^{ir}\\
	&=\left(\begin{array}{ccc}
0\\
 & 0\\
 &  & 1
\end{array}\right)\left(\begin{array}{c}
J_{x}^{ir}\\
J_{y}^{ir}\\
J_{z}^{ir}
\end{array}\right)
=J_{z}^{ir}[(2k)\omega_{0}]\boldsymbol{e}_{z}.
    \end{split}\label{J2k}
\end{align}
The prefactors ``4" in Jones matrices are absorbed into the Jones vectors. 
In the following derivation, the prefactors in Jones matrices are all ignored unless otherwise stated. 
Equations~\eqref{J2kp1} and \eqref{J2k} imply that the linearly polarized odd harmonics are permitted in the $x$ direction while the linearly polarized even harmonics are allowed in the $z$ direction.
This is consistent with our observations in the main text.

Then we consider the LP field deviates from high symmetry axes (like $x$ axis) shown in Fig.~\ref{laserjones}(b).
The shared symmetries are
\begin{align}
\mathcal{R}=\mathcal{C}_{2(z)}=\left(\begin{array}{ccc}
-1\\
 & -1\\
 &  & 1
\end{array}\right)
\text{ and }
     \mathcal{R}'=E=\left(\begin{array}{ccc}
1\\
 & 1\\
 &  & 1
\end{array}\right).\label{Rmy1}
\end{align}
Substituting Eq.~\eqref{Rmy1} into Eq.~\eqref{jms}, the Jones matrix is
\begin{align}
    \mathcal{J}(n_{H})=\left(\begin{array}{ccc}
1-e^{-i\pi n_{H}}\\
 & 1-e^{-i\pi n_{H}}\\
 &  & 1+e^{-i\pi n_{H}}
\end{array}\right). \label{c2jon2}
\end{align}
We thus find that the odd harmonics are allowed in both $x$ and $y$ directions, because the corresponding Jones matrix reads
\begin{align}
    \mathcal{J}(n_{H}=2k+1)=\left(\begin{array}{ccc}
1\\
 & 1\\
 &  & 0
\end{array}\right).
\end{align}
The odd harmonic field is expressed by
\begin{align}
    \begin{split}
        \boldsymbol{J}[(2k+1)\omega_{0}]	&=\mathcal{J}(2k+1)\boldsymbol{J}^{ir}\\
	&=\left(\begin{array}{ccc}
1\\
 & 1\\
 &  & 0
\end{array}\right)\left(\begin{array}{c}
J_{x}^{ir}\\
J_{y}^{ir}\\
J_{z}^{ir}
\end{array}\right) \\
&=\left(\begin{array}{c}
J_{x}^{ir}\\
J_{y}^{ir}\\
0
\end{array}\right).
    \end{split}
\end{align}
Since $J_{x}^{ir}$ and $J_{y}^{ir}$ are generally nonzero, the odd harmonics are elliptically polarized. This is verified by Fig.~\ref{thetadepen}(d). We can observe the nonzero harmonic ellipticity when the LP laser deviates from high symmetry axes.
Nevertheless, the even harmonics is still linearly polarized, due to
\begin{align}
    \mathcal{J}(n_{H}=2k)=\left(\begin{array}{ccc}
0\\
 & 0\\
 &  & 1
\end{array}\right).
\end{align}

\subsubsection{LCP driver in x-y plane}
Then we analyze the TaAs interacting with LCP laser, as shown in Fig.~\ref{laserjones}(c).
The shared symmetries are
\begin{align}
\mathcal{R}=\mathcal{C}_{4(z)}=\left(\begin{array}{ccc}
0 & -1 & 0\\
1 & 0 & 0\\
0 & 0 & 1
\end{array}\right)
\text{ and }
     \mathcal{R}'=E=\left(\begin{array}{ccc}
1\\
 & 1\\
 &  & 1
\end{array}\right),\label{Rc4}
\end{align}
where $\mathcal{C}_{4(z)}(x,y,z)^{T}=(-y,x,z)^{T}$. Plugging Eq.~\eqref{Rc4} into Eq.~\eqref{jms}, the Jones matrix is
\begin{align}
    \mathcal{J}(n_H)=\left(\begin{array}{ccc}
1-e^{-i\pi n_{H}} & 2i\sin(\frac{\pi}{2}n_{H}) & 0\\
-2i\sin(\frac{\pi}{2}n_{H}) & 1-e^{-i\pi n_{H}} & 0\\
0 & 0 & (1+i^{n_{H}})[1+(-1)^{n_{H}}]
\end{array}\right).
\label{jonec4}
\end{align}
Therefore, the permitted harmonic orders are $n_H=4k+1$, $4k-1$, and $4k$, and the corresponding Jones matrices have the forms of
\begin{align}
    \mathcal{J}(n_{H}=4k\pm1)=\left(\begin{array}{ccc}
\frac{1}{2} & \pm\frac{i}{2} & 0\\
\mp\frac{i}{2} & \frac{1}{2} & 0\\
0 & 0 & 0
\end{array}\right) \label{eq:4kp1}
\end{align}
and
\begin{align}
    \mathcal{J}(n_{H}=4k)=\left(\begin{array}{ccc}
0\\
 & 0\\
 &  & 1
\end{array}\right). \label{eq:4k}
\end{align}
As we know, $\mathcal{J}(n_{H}=4k)$ is a linear polarizer which only allows ($4k$)th linearly polarized harmonics along the $z$ direction. The Jones matrix $\mathcal{J}(n_{H}=4k+1)$ is a left circular polarizer, while $\mathcal{J}(n_{H}=4k-1)$ represents a right circular polarizer. 
Equation~\eqref{eq:4kp1} results in left-handed circularly polarized ($4k+1$)th harmonics in the $x$-$y$ plane, because
\begin{align}
    \boldsymbol{J}[(4k+1)\omega_{0}]	=\mathcal{J}(4k+1)\boldsymbol{J}^{ir}
	=\frac{1}{2}\left(\begin{array}{c}
J_{x}^{ir}+iJ_{y}^{ir}\\
-iJ_{x}^{ir}+J_{y}^{ir}\\
0
\end{array}\right),
\end{align}
and
\begin{align}
    \frac{J_{y}(4k+1)}{J_{x}(4k+1)}=\frac{-iJ_{x}^{ir}+J_{y}^{ir}}{J_{x}^{ir}+iJ_{y}^{ir}}=-i.
\end{align}
Nevertheless, the ($4k-1$)th harmonics are right-handed circularly polarized, because
\begin{align}
    \boldsymbol{J}[(4k-1)\omega_{0}]	=\mathcal{J}(4k-1)\boldsymbol{J}^{ir}
	=\frac{1}{2}\left(\begin{array}{c}
J_{x}^{ir}-iJ_{y}^{ir}\\
iJ_{x}^{ir}+J_{y}^{ir}\\
0
\end{array}\right),
\end{align}
and
\begin{align}
    \frac{J_{y}(4k-1)}{J_{x}(4k-1)}=\frac{iJ_{x}^{ir}+J_{y}^{ir}}{J_{x}^{ir}-iJ_{y}^{ir}}=i.
\end{align}
These also agree with our observations in the main text.

\begin{figure}
	\centering
	\includegraphics[width=0.55\columnwidth]{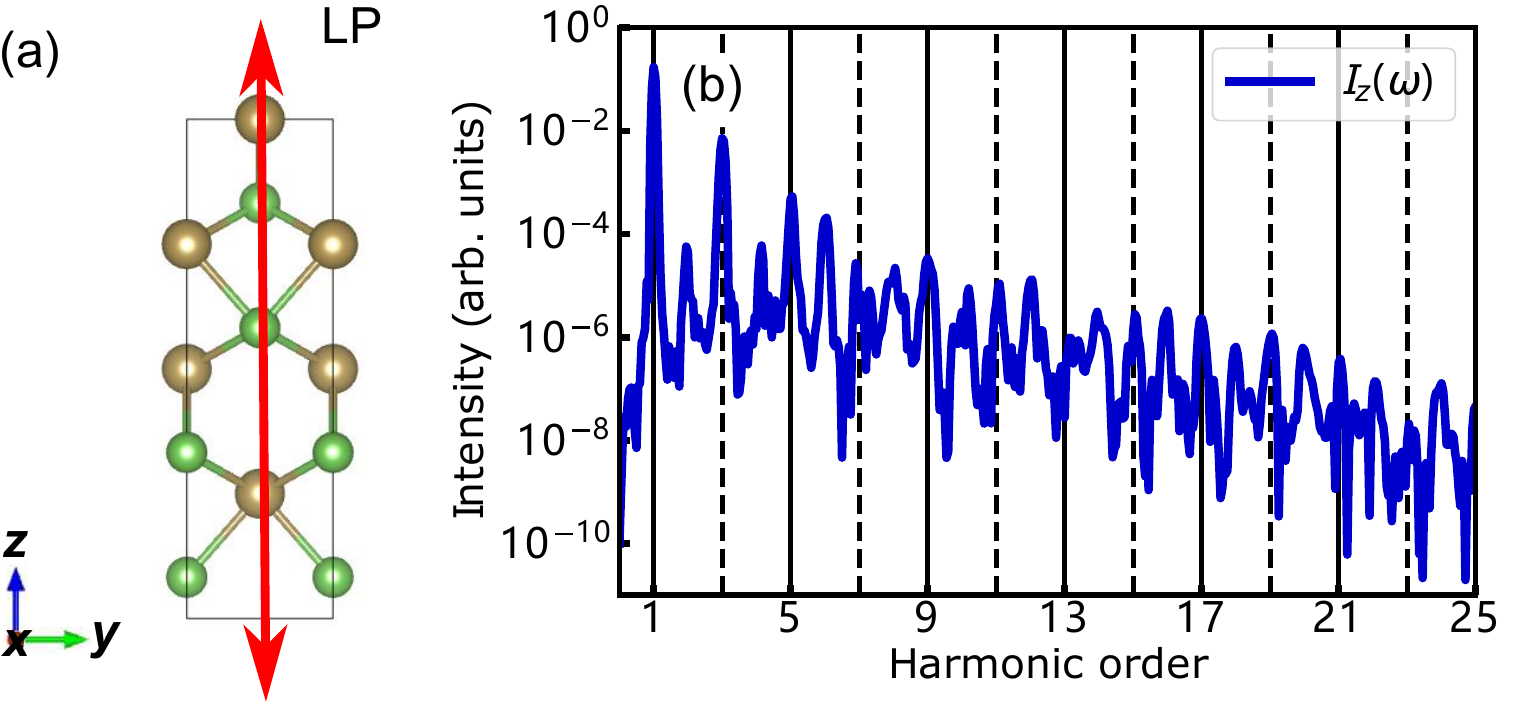}
	\caption{(a) Sketch of LP laser polarized along the $z$ axis of TaAs lattice. (b) Harmonic spectrum in the $z$ direction. The laser parameters are the same as those in Fig.~\ref{soc}.}
 \label{LPalongz}
\end{figure}

\subsubsection{LP driver along z axis}
As the LP laser polarized along the $z$ direction, 
we find that no crystal symmetry can result in a translation of the laser field in time. 
Thus the shared symmetry $\mathcal{R}$ is $E$. 
However, the LP laser is invariant under the operation of $\mathcal{M}_x$, $\mathcal{M}_y$, $\mathcal{M}_{xy}$, $\mathcal{M}_{-xy}$, $\mathcal{C}_{2(z)}$, $\mathcal{C}_{4(z)}$, and $\mathcal{C}_{4-(z)}$, see Fig.~\ref{LPalongz}(a). Plugging $\mathcal{R}=E$ and $\mathcal{R}'=\mathcal{M}_x,\mathcal{M}_y$ into Eq.~\eqref{jms}, the Jones matrix is expressed as
\begin{align}
    \mathcal{J}(n_{H}=k)=\left(\begin{array}{ccc}
0\\
 & 0\\
 &  & 1
\end{array}\right). \label{JnHk}
\end{align}
Equation~\eqref{JnHk} indicates that both odd and even harmonics are permitted in the $z$ direction, while harmonic emissions are prohibited in other directions. This is verified by the HHG spectrum shown in Fig.~\ref{LPalongz}(b).

\begin{figure}
	\includegraphics[width=0.9\columnwidth]{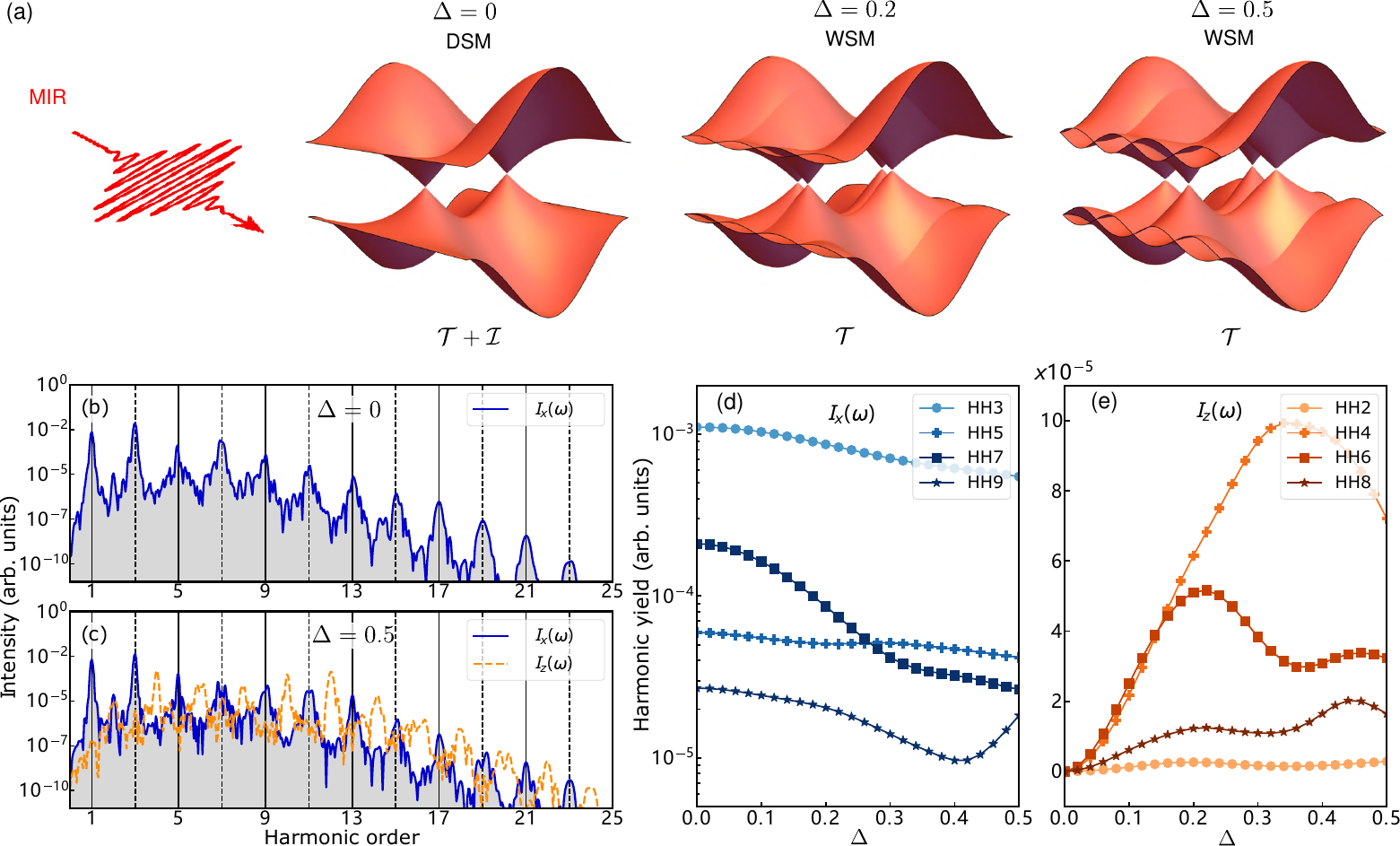}
	\caption{(a) Band structure of minimal model for TaAs at $k_z=0$ plane with different inversion symmetry parameters. HHG spectra for (b) $\Delta=0$ and (c) $\Delta=0.5$ driven by a linearly polarized field. Panels (d) and (e) show the yields of even- and odd-order harmonics as a function of inversion symmetry parameter $\Delta$, respectively. Hamiltonian parameters are set to be $t=0.8$ eV, $m_y=1.0$, $m_z=3.0$, and $a=3.44$ {\AA}. The laser intensity is $I_{0} = 5.6 \times 10^{11}$ W/cm$^2$ ($E_{0} = 0.004$ a.u.), and other laser parameters are the same as those in Fig.~\ref{soc}.}
	\label{tbmodel}
\end{figure}

\section{\label{sec:invers}Role of inversion symmetry breaking}

In order to clarify the impact of $\mathcal{I}$ breaking on HHG from TaAs, we utilize a four-band minimal model ($\mathcal{T}$ is preserved) \cite{Wu2017NLO}
\begin{align}
    \begin{split}\label{tbTaAs}
        H_{0}(\boldsymbol{k})=&t\{\cos(k_{x}a)+m_{y}[1-\cos(k_{y}a)]+m_{z}[1-\cos(k_{z}a)]\}\sigma_{x}\\
    &+t[\sin(k_{y}a)+\Delta\cos(k_{y}a)s_{x}]\sigma_{y} +t\sin(k_{z}a)s_{x}\sigma_{z},
    \end{split}
\end{align}
which supports four WPs in $k_z=0$ plane capturing main physics near the Fermi energy. Here, $\sigma_i$ and $s_i$ denote Pauli matrices symbolizing the orbital and spin degrees of freedom, respectively. $t$ is the hopping strength, 
$a$ is the lattice constant, $m_{y}$ and $m_{z}$ are parameters that 
introduce anisotropy, and $\Delta$ introduces $\mathcal{I}$ breaking. Roughly speaking, the larger the $\Delta$, the greater the inversion symmetry is broken. This effective model preserves the $mm2$ point group, which is
a subgroup of $4mm$ point group of crystal TaAs.

Firstly, we discuss the crystal symmetries in the model. For $\Delta=0$, the model has $\mathcal{I}$, $\mathcal{M}_{x}$, $\mathcal{M}_{y}$, $\mathcal{M}_{z}$, $\mathcal{C}_{2(x)}$, $\mathcal{C}_{2(y)}$, and $\mathcal{C}_{2(z)}$ symmetries. 
It is worth to note that these symmetries are not independent, because $\mathcal{I}=\mathcal{M}_{x}\mathcal{M}_{y}\mathcal{M}_{z}=\mathcal{M}_{z}\mathcal{C}_{2(z)}=\mathcal{M}_{x}\mathcal{C}_{2(x)}=\mathcal{M}_{y}\mathcal{C}_{2(y)}$.
For $\Delta\neq 0$, the $\Delta$ term breaks the $\mathcal{M}_{z}$ symmetry, which naturally results in the absence of  $\mathcal{I}$, $\mathcal{C}_{2(x)}$, and $\mathcal{C}_{2(y)}$. 
Only $\mathcal{M}_{x}$, $\mathcal{M}_{y}$, and $\mathcal{C}_{2(z)}$ symmetries are preserved. The broken $\mathcal{M}_{z}$ is exactly the reason causes $\mathcal{I}$ breaking in the realistic TaAs, as shown in Fig.~\ref{LPalongz}(a).

Secondly, we focus on the HHG in the TaAs model. In the scenario where $\Delta=0$, the exist of both $\mathcal{T}$ and $\mathcal{I}$ symmetries leads to a Dirac semimetal (DSM) phase.
Two Dirac nodes emerge at $\boldsymbol{k}= (\pm \pi/2a, 0,0)$ as illustrated in Fig.~\ref{tbmodel}(a).
Consequently, only odd harmonics are observable in the $x$ direction
when a LP pulse is applied along the $x$ direction, as depicted in Fig.~\ref{tbmodel}(b).
With increasing $\Delta$ (breaking $\mathcal{M}_{z}$), 
the Dirac points split into pairs of WPs with opposite chirality and displace in opposite directions along the $k_y$ axis ($\Delta k_y\propto\Delta/a$). 
In the case of $\Delta=0.5$, as shown in Fig.~\ref{tbmodel}(c),  even harmonics appear in the $z$ direction because of $\mathcal{C}_{2(z)}$ symmetry. 
In Figs.~\ref{tbmodel}(d) and \ref{tbmodel}(e), we present how the yield of ordinary and anomalous harmonics varies with increasing the inversion breaking parameter $\Delta$. 
Within the range ($\Delta<0.2$) where the $\mathcal{I}$ breaking term can be treated as a perturbation to the Hamiltonian, the yields of 
odd harmonics are suppressed while the yields of even harmonics increase monotonously. 
However, the variation of harmonic yields deviates from a monotonic pattern, as $\Delta$ enters nonperturbation zone ($\Delta>0.2$). 

Finally, the Jones matrix formalism suggests that it is necessary to identify which sub-symmetry absence leads to the breaking of $\mathcal{I}$ when focusing on the selection rule of HHG. 
In our case, the absence of $\mathcal{M}_{z}$ results in odd harmonics in the $x$ direction but even harmonics in the $z$ direction when the laser is polarized along the $x$ direction. However, both odd and even harmonics can be observed in the $x$ and $y$ directions in other systems if the broken $\mathcal{I}$ stems from the absence of $\mathcal{C}_{2(z)}$. In the $z$ direction, harmonic emission would not be allowed.

\section{\label{sec:orenap}Orientation dependence}
\begin{figure}
	\centering
	\includegraphics[width=0.4\columnwidth]{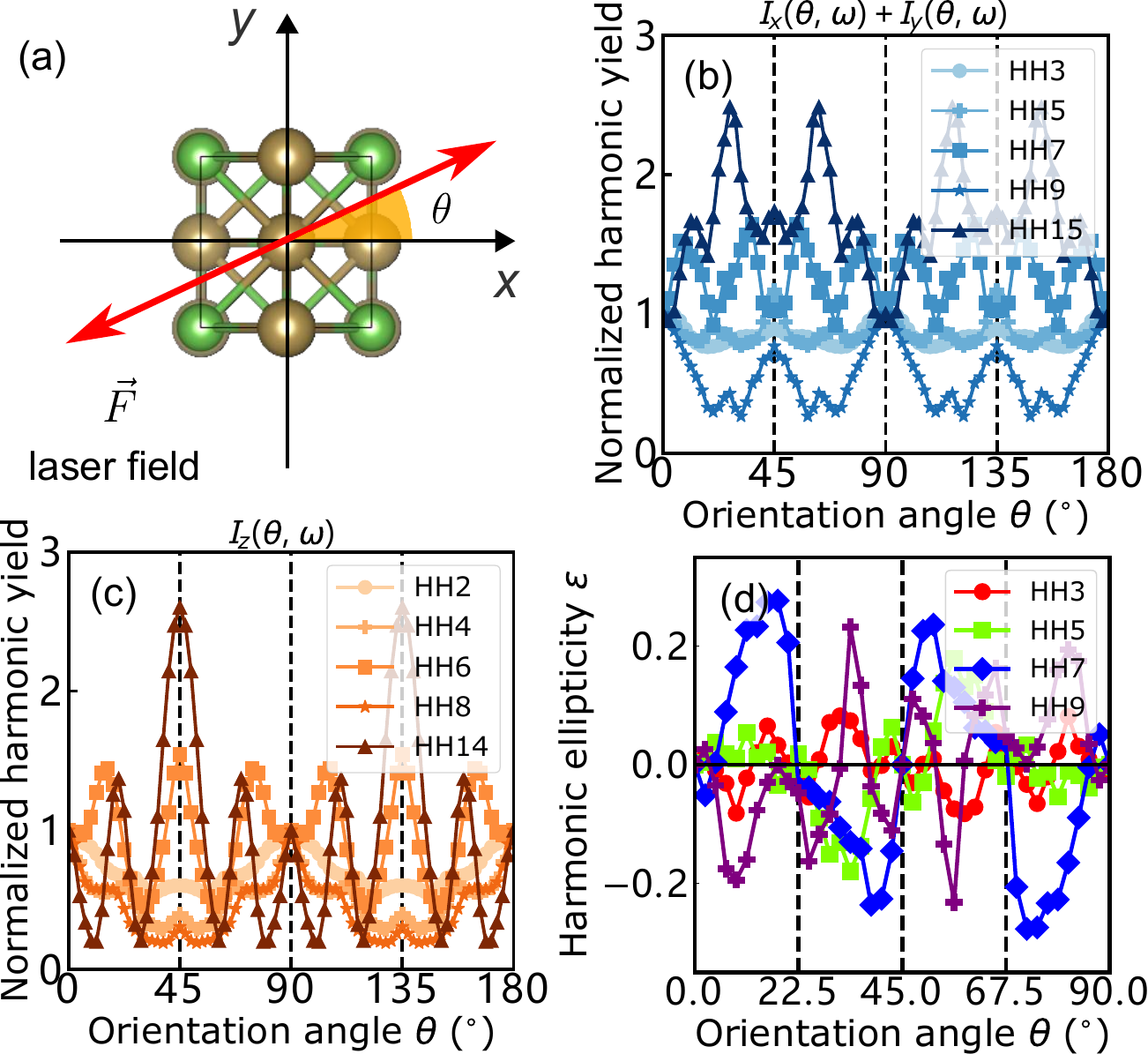}
	\caption{(a) Sketch of the LP field and crystal structure, where $\theta$ is the angle of electric field relative to the $x$ axis. (b) and (c) Normalized yields of the even- and odd-order harmonics vs orientation angle $\theta$. (d) Ellipticity of emitted odd-order harmonics vs orientation angle. The laser parameters are the same as those in Fig.~\ref{soc}.}
	\label{thetadepen}
\end{figure}

In Figs.~\ref{thetadepen}(b) and \ref{thetadepen}(c), we find that the crystal symmetries of TaAs are reflected in the angular dependence of both ordinary ($x$-$y$ plane) and anomalous harmonics ($z$ direction).
The $\mathcal{C}_{4(z)}$ and $\mathcal{C}_{4-(z)}$ strongly modulate the harmonic yields with a common period of $90 \degree $. 
The mirror symmetries ($\mathcal{M}_{x}$, $\mathcal{M}_{y}$, $\mathcal{M}_{xy}$, 
and $\mathcal{M}_{-xy}$) ensure the symmetric profiles of harmonic yields concerning the axes at $45\degree k$, where $k \in \mathbb{N}$.
The odd harmonics are elliptically polarized, when the LP laser deviates from high symmetry axes ($\theta = 45\degree k$) as shown in
Fig.~\ref{thetadepen}(d). 
It has been explained by Jones matrix formalism in Sec.~\ref{sec:EJMa}.

\section{\label{sec:CDich}Circular dichroism}
\begin{figure}
	\centering
	\includegraphics[width=0.8\columnwidth]{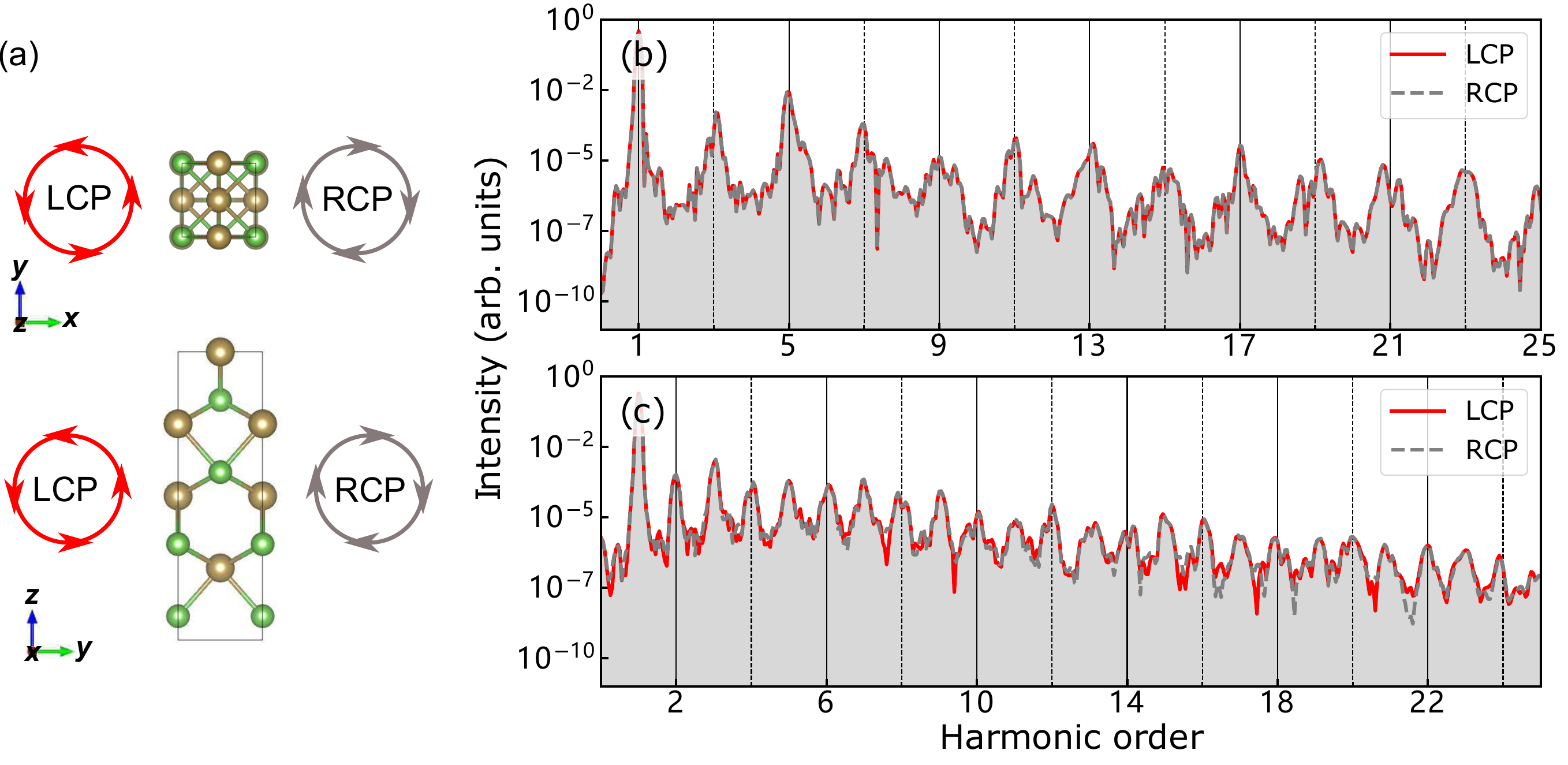}
	\caption{(a) TaAs lattice driven by LCP and RCP fields in the $x$-$y$ and $y$-$z$ planes. HHG spectra excited by LCP and RCP lights, where polarization planes are (b) $x$-$y$ and (c) $y$-$z$, respectively. }
 \label{LRCP}
\end{figure}
As shown in Fig.~\ref{LRCP}(a), we use circularly polarized lights with opposite helicities to excite the TaAs crystal on the $x$-$y$ and $y$-$z$ planes respectively. The HHG excited by circular polarized field on the $x$-$z$ plane is the same as that excited by the field on the $y$-$z$ plane, because of $\mathcal{C}_{4(z)}$ symmetry.
In Figs.~\ref{LRCP}(b)-(c), we find that the HHG spectra excited by LCP light are identical to those excited by RCP light. Moreover, using Eq.~\eqref{CDn}, we obtain the HHG circular dichroism of each order is zero.

\section{\label{sec:Ellip}Ellipticity dependence}
\begin{figure}
	\centering
	\includegraphics[width=0.4\columnwidth]{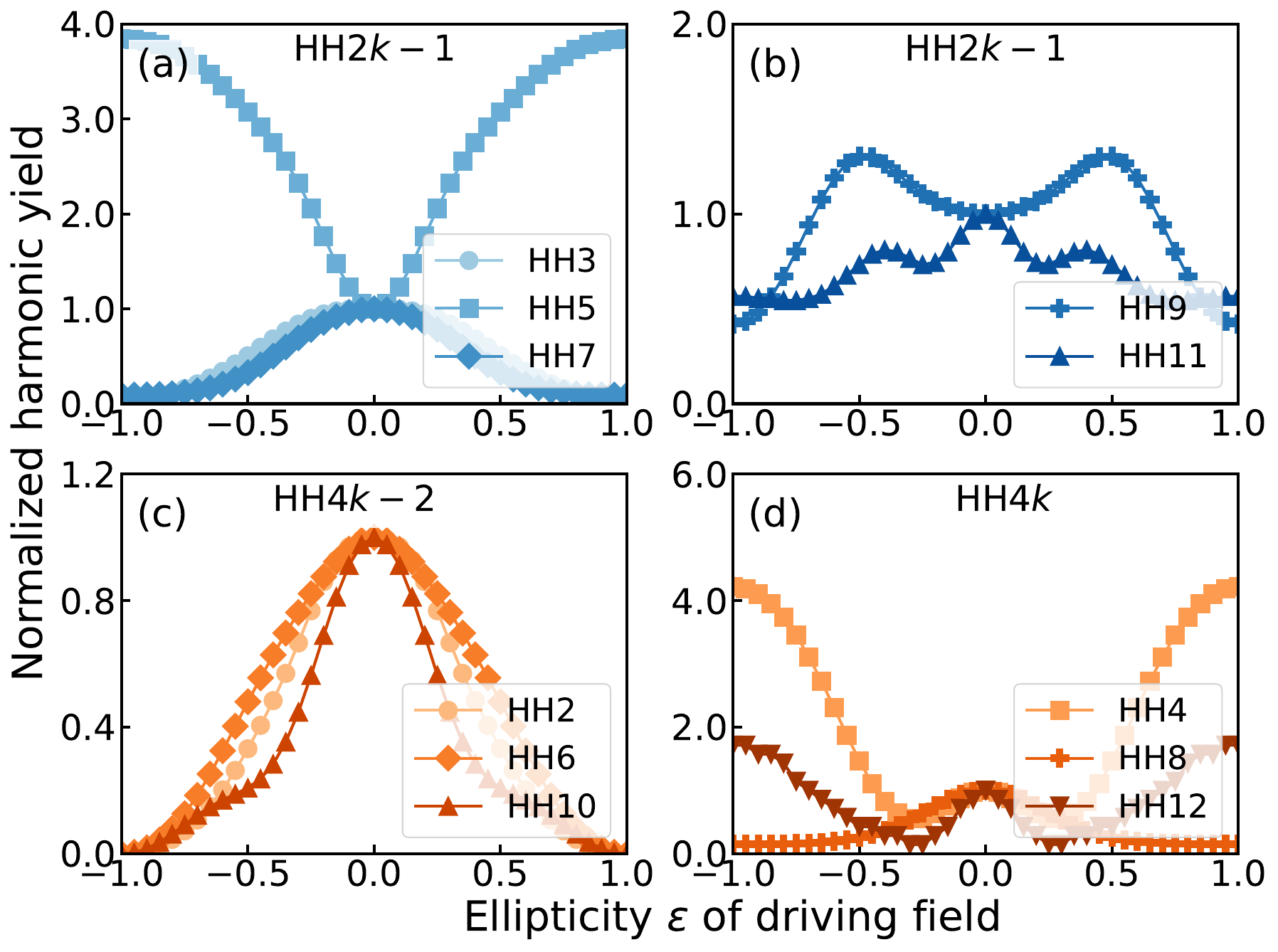}
	\caption{Ellipticity dependence of HHG in TaAs driven by a weaker laser field. The intensity is $I_{0} = 5.6 \times 10^{9}$ W/cm$^2$ ($E_{0} = 0.0004$ a.u.), and the wavelength is $\lambda=3200$ nm ($\hbar\omega=0.388$ eV).}
 \label{EDweak}
\end{figure}

In TaAs, we find that the ellipticity dependence of HHG is sensitive to the parameters of driving field. In comparison to the strong laser field in the main text, the laser field with lower intensity and shorter wavelength gives rise to a relatively normal behavior of HHG ellipticity dependence, illustrated in Fig.~\ref{EDweak}. The yields of 2th, 3th, 6th, 7th, 8th, and 10th harmonics monotonically decay with increasing the driver ellipticity.

\begin{figure}
	\centering
	\includegraphics[width=\columnwidth]{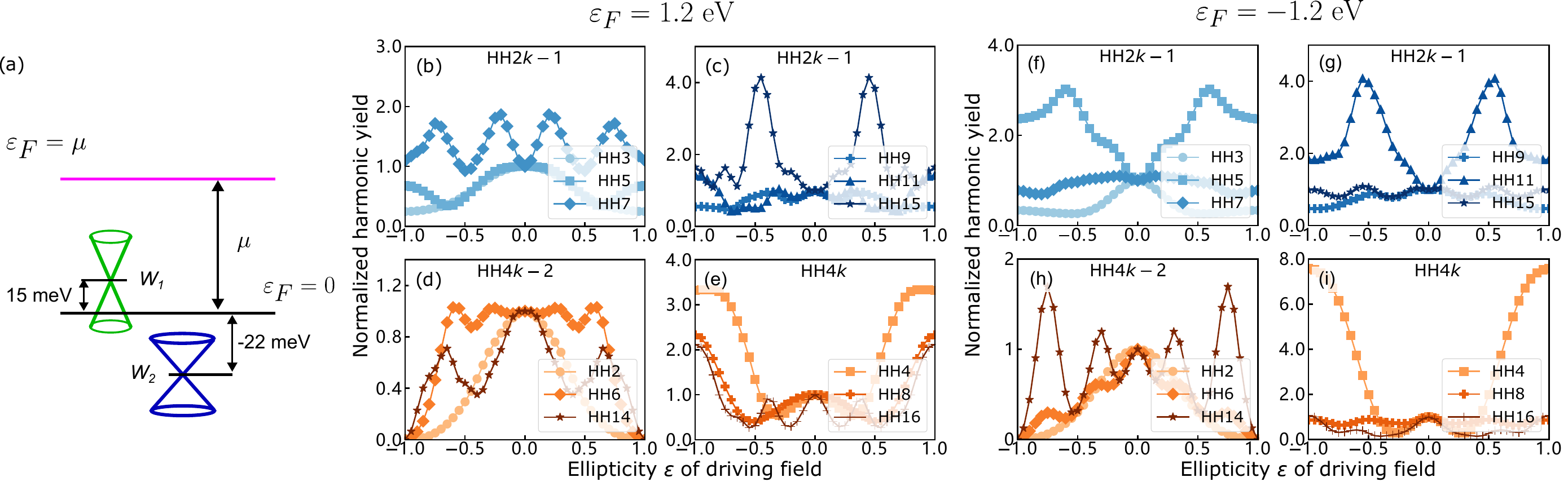}
	\caption{(a) Relative positions of Weyl cones ($W_1$ and $W_2$) and Fermi energy. Fermi energy is shifted by tuning chemical potential $\mu$. Ellipticity dependence of HHG in TaAs with (b)-(e) $\varepsilon_{F}=1.2$ eV and (f)-(i) $\varepsilon_{F}=-1.2$ eV. The laser parameters are the same as those in Fig.~\ref{soc}.}
 \label{EDfermi}
\end{figure}

Unlike graphene HHG, HHG in TaAs consistently exhibits anomalous behavior, even when the Fermi energy is tuned to be significantly higher ($\varepsilon_F=1.2$ eV) or lower ($\varepsilon_F=-1.2$ eV) than the energy of the Weyl nodes as shown in Fig.~\ref{EDfermi}.

%